\documentclass[intlimits,twoside,a4paper]{article}

\usepackage[cp1251]{inputenc}
\usepackage[eqsecnum]{cmpj3}

\usepackage{xcolor}

\articletype{Review}

\issue{2023}{26}{2}{22701}
\doinumber{10.5488/CMP.26.22701}
\title[Structure and properties of the films]
{Structure and properties of the films based on ternary transition metal
borides: \\ theory and experiment}
\author[A. A. Onoprienko, V. I. Ivashchenko, V. I. Shevchenko]%
{A. A. Onoprienko\orcid{0000-0001-7173-0404}, V. I. Ivashchenko\orcid{0000-0002-1439-8449}, V. I. Shevchenko\orcid{0000-0001-8427-4751}%
\thanks{Corresponding author: \email{shev@ipms.kiev.ua}.}}
\address{Frantsevich Institute for Problems of Materials Sciences,
NAS of Ukraine, Kyiv, Ukraine}


\date{Received October 17, 2022, in final form January 31, 2023}

\begin{document}

\maketitle

\begin{abstract}
  The review presents the results of theoretical and experimental studies
  of the structure, bonding between atoms, mechanical properties, thermal
  stability, and oxidation and corrosion resistance of films based on
  ternary transition metal borides.
  \printkeywords{}
\end{abstract}

\section{Introduction}

The increasing industrial demand for protective coatings with high hardness,
good elastic properties and thermal stability calls for the investigation of
new materials.
Though transition metal nitrides and carbides are successfully used for
different tasks in automotive or aerospace industries, the search for
improved materials is an essential problem.

Boron is one of the hardest materials known.
Therefore, a promising way to achieve strong materials with exceptional
properties is the investigation of borides, in particular of
\textcolor{black}{transition metal (TM)} borides.
Transition metal borides showed potential applications as hard protective
thin films and electrical contact materials.
Among the stoichiometric variety of transition metal borides (TM-B, TM-B$_2$,
TM-B$_4$, TM-B$_{14}$, etc.), the diborides exhibit useful properties such as
high hardness and wear resistance, high melting points, high conductivity,
chemical inertness, good corrosion resistance, and good thermal
stability~\cite{1,2,3,4,5,6,7,8,9,10,11,12,13,14,15}.

\begin{figure}[htpb]
  \centering
  \includegraphics[width=0.55\linewidth]{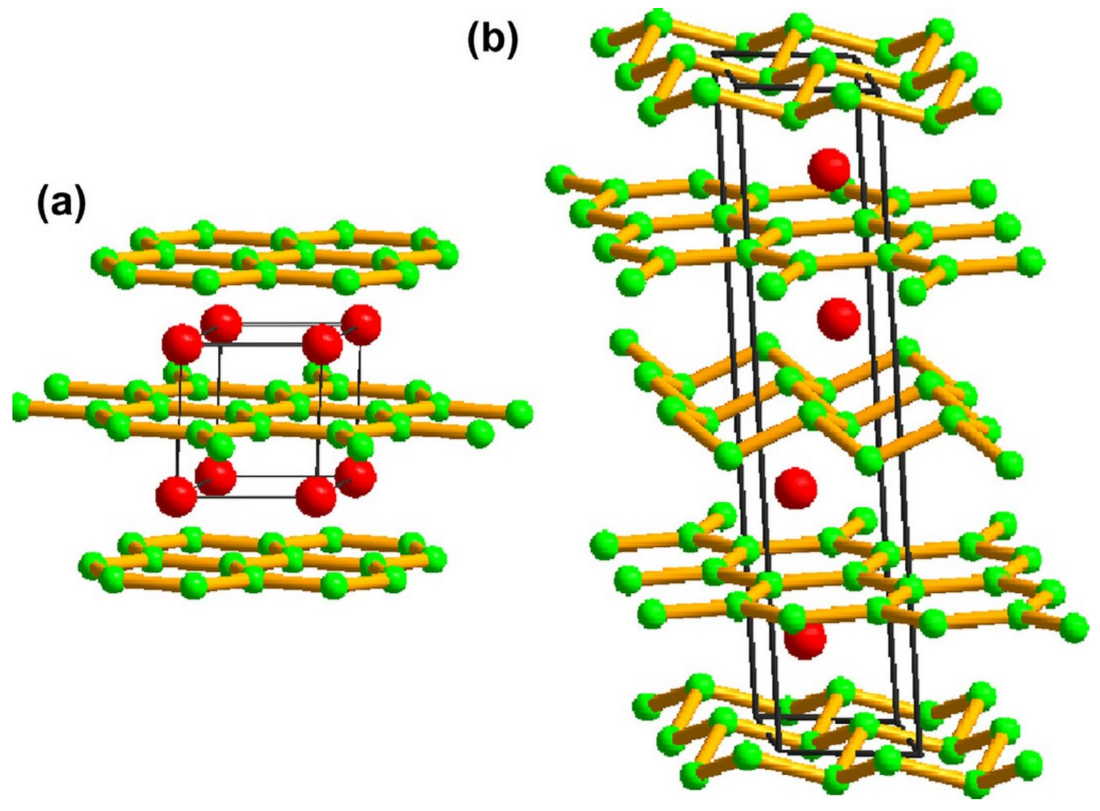}
  \caption{\label{fig:1}(Colour online)
    Unit cell and layer structure of (a) AlB$_2$ and (b) WB$_2$
    prototypes~\cite{16}.
  }
\end{figure}

The binary TM-B$_2$ compounds are known to crystallize in two related
hexagonal structures: $\alpha$-AlB$_2$-prototype or
$\omega$-W$_2$B$_{5-z}$-prototype (figure~\ref{fig:1})~\cite{16}.
A large number of early TMB$_2$ compounds (such as TiB$_2$, ZrB$_2$, VB$_2$,
etc.) have the AlB$_2$ structure type and crystallize in a three atoms unit
cell with space group 191 ($P6/$mmm).
An instructive description of this structure as a stacking of hexagonal
planes with covalently bonded boron atoms that are separated by metal layers
is given in figure~\ref{fig:1}(a).
The boron atoms form graphite-like covalently bonded hexagons, with metal
atoms above (and below) their centers.
In addition to the predominant AlB$_2$ structure type, different structural
modifications of diboride phases are known peculiar to late TMB$_2$ (such as
WB$_2$, ReB$_2$, TaB$_2$, etc.).
One such example is WB$_2$, for which different modifications are known.
Recently, WB$_2$ thin films were reported to crystallize in the AlB$_2$
structure, whereas bulk material prefers the WB$_2$ modification, formerly
known as W$_2$B$_5$.
The WB$_2$ structure type is closely related to the AlB$_2$ prototype but it
evidences a different layer structure, with both flat and puckered boron
layers~\cite{16}.
This results in an increased unit cell, containing twelve atoms and
crystallizing in space group 194 ($P63$/mmc) as depicted in
figure~\ref{fig:1}~(b).

Ternary transition metal borides are potential candidates as a coating for
wear resistant applications when combining different structural modifications
of TM-diborides (AlB$_2$- and WB$_2$-structured) which exhibit good
exploitation properties.
However, while ternary transition metal nitrides and carbides as coatings
were investigated in detail~\cite{17}, ternary transition metal boride
coatings are still much less explored.
There is a limited amount of papers in which the ternary transition metal
borides \textcolor{black}{(M1,M2)B},
where M1, M2 = La, Zr, W, Ti, Ta, Fe, Cr, Ni, Sc, Hf, V, Nb, Co, Mo, are
studied.

\textcolor{black}{On the other hand, first principles calculations of the
  structure and properties of such ternary borides are very important since
  the search of the optimum deposition parameters by means of trial and error
  method is not effective.
  A more promising approach is to combine experiment with calculations.
  Therefore, in our review we paid much attention to the calculations of
  the structure and mechanical properties of ternary boride systems.
  So far, there are no comprehensive review papers on the films based on
  ternary boride systems in which the first principles investigations could be
  reviewed together with experimental studies.
  Thus, in this paper we aimed at filling this gap in studying such a kind of
films.}

\section{Theoretical studies}

\textcolor{black}{In combination with experimental research, the computational
  materials science creates an effective instrument for forecasting the
  structure and properties of materials.
  The structure and properties of some ternary transition metal borides were theoretically studied based on the first principle calculations within
the framework of density functional theory (DFT).}

As it was noted in Introduction section, the transition metal diborides can
exist in two structural prototypes, namely: AlB$_2$ and WB$_2$.
Despite the fact that the AlB$_2$ structure type is the predominant one,
there also exist diboride phases which prefer to crystallize in other
modifications.
One such phase is WB$_2$, for which the two above mentioned structural modifications are
reported.
Due to the existence of the different structural modifications, combining
AlB$_2$-structured TM-diborides with WB$_2$ ones can result in ternary
systems that are based on combined allotropes.
Euchner et al.~\cite{16} using DFT calculations investigated the supersaturated
solid solution of ternary systems Ti$_x$W$_{1-x}$B$_2$ and
V$_x$W$_{1-x}$B$_2$ with respect to structure formation and stability.
To distinguish structural modifications, the symbols ``a'' and ``w'' were used to
represent the AlB$_2$ (a-M$_x$W$_{1-x}$B$_2$) and the WB$_2$ prototypes
(w-M$_x$W$_{1-x}$B$_2$), respectively.

\begin{figure}[htpb]
  \centering
  \includegraphics[width=0.4\linewidth]{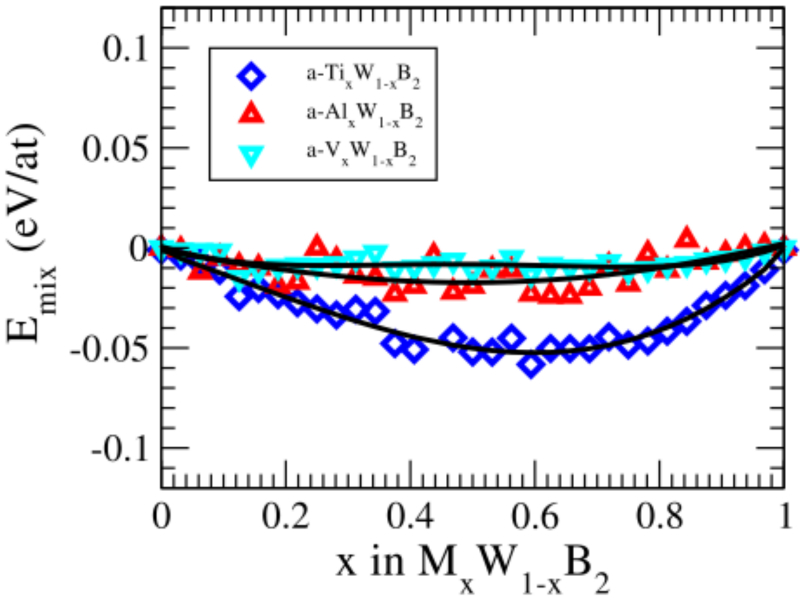}
  \caption{(Colour online) Enthalpy of mixing of M$_x$W$_{1-x}$B$_2$ with respect to the
    a-phase.
  The black curves are fits to the data~\cite{16}.}
  \label{fig:02}
\end{figure}

Supersaturated solid solutions of a-M$_x$W$_{1-x}$B$_2$ are highly stable
since they essentially show no tendencies for spinodal decomposition.
When plotting the differences in energy of formation between
a-M$_x$W$_{1-x}$B$_2$ and the isostructural constituent phases a-AlB$_2$ and
a-WB$_2$, it becomes evident that the a-phase is stable against isostructural
decomposition since $\Delta E_\mathrm{mix}$ is negative (or very close to
zero) for the whole composition range (cf. figure~\ref{fig:02}).
This will be even further enhanced when configurational entropy is taken into
account.

The fact that an interaction between two different allotropes may result in
materials with improved physical properties is well known from the
Ti$_{1-x}$Al$_x$N system.
These improvements in the physical properties originate from the interaction
between the preference for hexagonal and cubic structure types of AlN and
TiN.
Following this idea of competing structure types, supersaturated solid
solutions of ternary borides were studied, which are based on binary
constituents that in principle prefer to crystallize in different
modifications.
It was shown on the exemplary cases of Ti$_{x}$W$_{1-x}$B$_2$ and
V$_x$W$_{1-x}$B$_2$ that such ternary borides represent a new class of
metastable materials which offer a large field for further investigations.
The recent successful deposition of a-WB$_2$ together with the calculated
formation energies allows the authors of \cite{16} to conclude that
solid solutions of a-M$_x$W$_{1-x}$B$_2$ type are experimentally accessible
over a large composition range.

In \cite{18}, the crystal structures, mechanical properties and Debye
temperatures of 18 transition metal ternary borides
M$^\mathrm{I}_x$M$^\mathrm{II}$B$_x$ ($x$ = 1 or 2, M$^\mathrm{I}$ and
M$^\mathrm{II}$ are transition metals) were studied through
first-principle calculations.
The thermodynamic stability of the ternary borides crystal structures was
estimated by the cohesive energy ($E_\mathrm{coh}$) and formation enthalpy
($\Delta H_\mathrm{r}$).
When both $E_\mathrm{coh}$ and $\Delta H_\mathrm{r}$ are negative, the
crystal structure is thermodynamically stable and the smaller the negative
values are, the more stable the structure is.

The mechanical stability of the M$^\mathrm{I}_x$M$^\mathrm{II}$B$_x$
compounds was estimated using the criteria for elastic constants $c_{ij}$.
The bulk modulus $B$, shear modulus $G$, Young's modulus $E$, and Poisson's
ratio $\nu$ of the compounds were also determined.
The ductility of a material can be evaluated by the $B/G$ ratio and Poisson's
ratio $\nu$, namely, the larger the $B/G$ and $\nu$ values are, the better is the
ductility.
Hardness is a key indicator for predicting the wear resistance of coating
materials.
Debye temperature is closely related to solid lattice vibration, thermal
conductivity, thermal expansion and specific heat: solid materials with
higher Debye temperatures tend to exhibit higher thermal conductivity.

\begin{figure}[!t]
  \centering
  \includegraphics[width=0.8\linewidth]{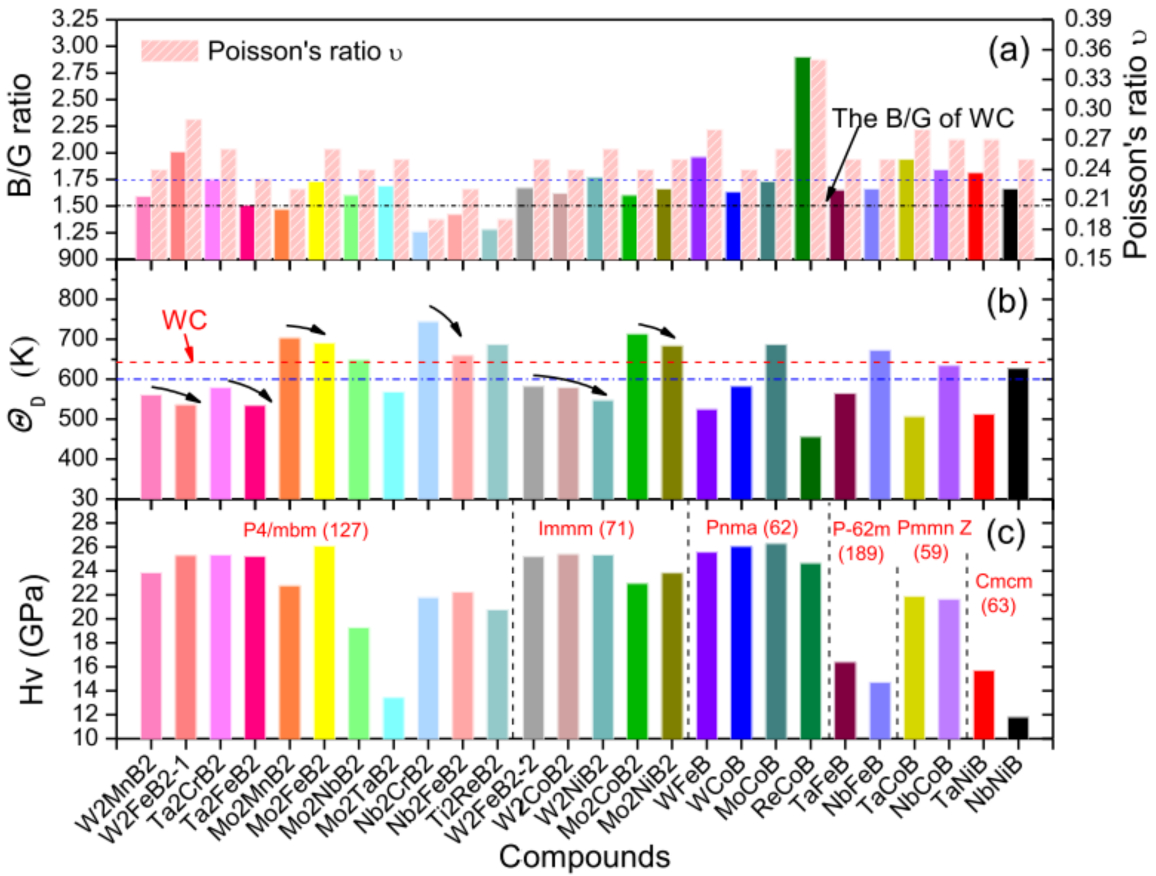}
  \caption{(Colour online) Some features of the  M$^\mathrm{I}_x$M$^\mathrm{II}$B$_x$
    ($x$ = 1 or 2):
    (a) $B/G$ ratio and Poisson's ratio $\nu$;
    (b) Debye temperature;
    (c) Hardness.
    For interpretation of the references to colour in this figure legend,
  the reader is referred to the Web version of this article \cite{18}.}
  \label{fig:3}
\end{figure}

The calculated Vickers hardness, Debye temperature, $B/G$ ratio and Poisson's
ratio are shown in figure~\ref{fig:3}.
The main conclusions following from figure~\ref{fig:3} are as follows:
(1) the calculated formation enthalpy and respective elastic constants
indicate that all the phases possess thermodynamic stability and mechanical
stability;
(2) the results of $B/G$ values confirm that the
M$^\mathrm{I}_x$M$^\mathrm{II}$B$_x$ ($x$ = 1 or 2) except Ta$_2$FeB$_2$,
Mo$_2$MnB$_2$, Nb$_2$CrB$_2$, Nb$_2$FeB$_2$ and Ti$_2$ReB$_2$, had a better
ductility than the compound WC; besides, W$_2$MnB$_2$, Mo$_2$FeB$_2$,
Mo$_2$NbB$_2$, Nb$_2$CrB$_2$, Nb$_2$FeB$_2$, Ti$_2$ReB$_2$, Mo$_2$CoB$_2$,
Mo$_2$NiB$_2$, MoCoB and NbFeB show a larger Debye temperature than WC;
(3) the predicted hardness of the M$^\mathrm{I}_x$M$^\mathrm{II}$B$_x$ ($x$~=~1 or 2) ternary borides, based on the calculated overlap populations and Gao's
hardness model, show that MoCoB exhibits the maximum hardness value of 26.3
GPa, which is still lower than that of WC; (4)~by analyzing the density of states and Poisson's ratio $\nu$, all the
M$^\mathrm{I}_x$M$^\mathrm{II}$B$_x$ ($x$ = 1 or 2) exhibit metallic, ionic
and covalent hybrid properties;
(5) according to the calculated results and phase diagrams, a high hardness
and thermal conductivity of MoCoB-Co cermet consisting of hard phase MoCoB
particles in a ductile Co matrix is proposed as a material with a particular
promise for replacing WC in some applications.
Obviously, all the hardness values of M$^\mathrm{I}_x$M$^\mathrm{II}$B$_x$
($x$ = 1 or 2) are less than the hardness of tungsten carbide (27--33.3 GPa),
which means that there is still a certain difference between the ternary
boride and the tungsten carbide considering only the hardness of the hard
phase.
However, these gaps are not large.
In fact, the hardness of a material is not only affected by the intrinsic
factors such as chemical composition and the underlying arrangement of atoms,
but also by the extrinsic effects such as surface morphology and grain
structure of the material.

The results obtained  also showed that the $E_\mathrm{coh}$ and
$\Delta H_\mathrm{r}$ of all the ternary transition metal borides are
negative, indicating that the structures are stable.
All the calculated elastic constants $c_{ij}$ indicated a structural stability
of the ternary borides considered.
The ternary transition metal borides studied exhibited the hardness in the range
of $\sim$12 GPa (NbNiB) to $\sim$26 GPa (MoCoB), high ductility and metallic,
ionic and covalent hybrid properties.

Alling et al.~\cite{19} carried out investigations by first-principles using the density functional theory of the mixing thermodynamics of
the alloy systems formed by M$^\mathrm{I}_{1-x}$M$^\mathrm{II}_x$B$_x$
combinations of the diborides ScB$_2$, YB$_2$, TiB$_2$, ZrB$_2$, HfB$_2$,
VB$_2$, NbB$_2$, and TaB$_2$, all reported to crystallize in the AlB$_2$ type
structure.

\begin{figure}[htpb]
  \centering
  \includegraphics[width=0.7\linewidth]{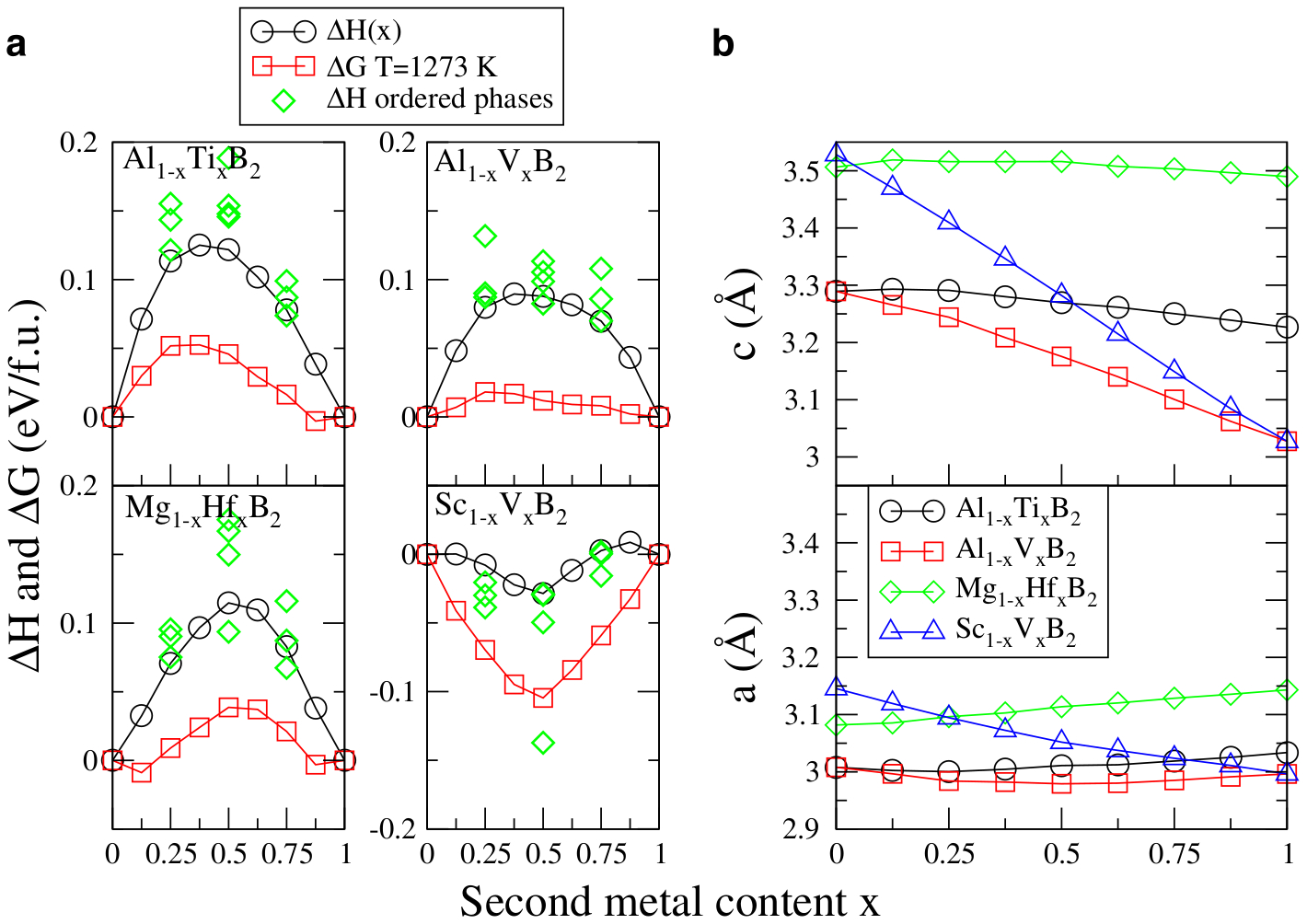}
  \caption{(Colour online) a) Mixing enthalpies of disordered and ordered alloys, as well as the
    mean field free energy at 1000\textcelsius{} for alloys that are identified
    as lattice matched clustering candidates, and an example of an ordering
    system, and
  b) $c$- and $a$-lattice parameters as a function of composition~\cite{19}.}
  \label{fig:4}
\end{figure}

It is well known that lattice-matched systems can display clustering.
Such alloys, with a small lattice mismatch, but a strong driving force for
clustering are of particular interest for age-hardening potential.
They are likely to form fully coherent interfaces and display spinodal
decomposition with large composition fluctuations even when diffusion is
limited.
The resulting nanostructure in the lattice can decrease dislocation mobility
and increase hardness.
For this reason, the alloys Al$_{1-x}$Ti$_x$B$_2$, Al$_{1-x}$V$_x$B$_2$,
Mg$_{1-x}$Hf$_x$B$_2$ deserve further attention because they are predicted to
display clustering and a small volume misfit.
Figure~\ref{fig:4} shows \textcolor{black}{the mixing enthalpies} of ordered and
disordered alloys for these systems, as well as for the ordering
Sc$_{1-x}$V$_x$B$_2$ system for comparison.

\begin{figure}[htpb]
  \centering
  \includegraphics[width=0.4\linewidth]{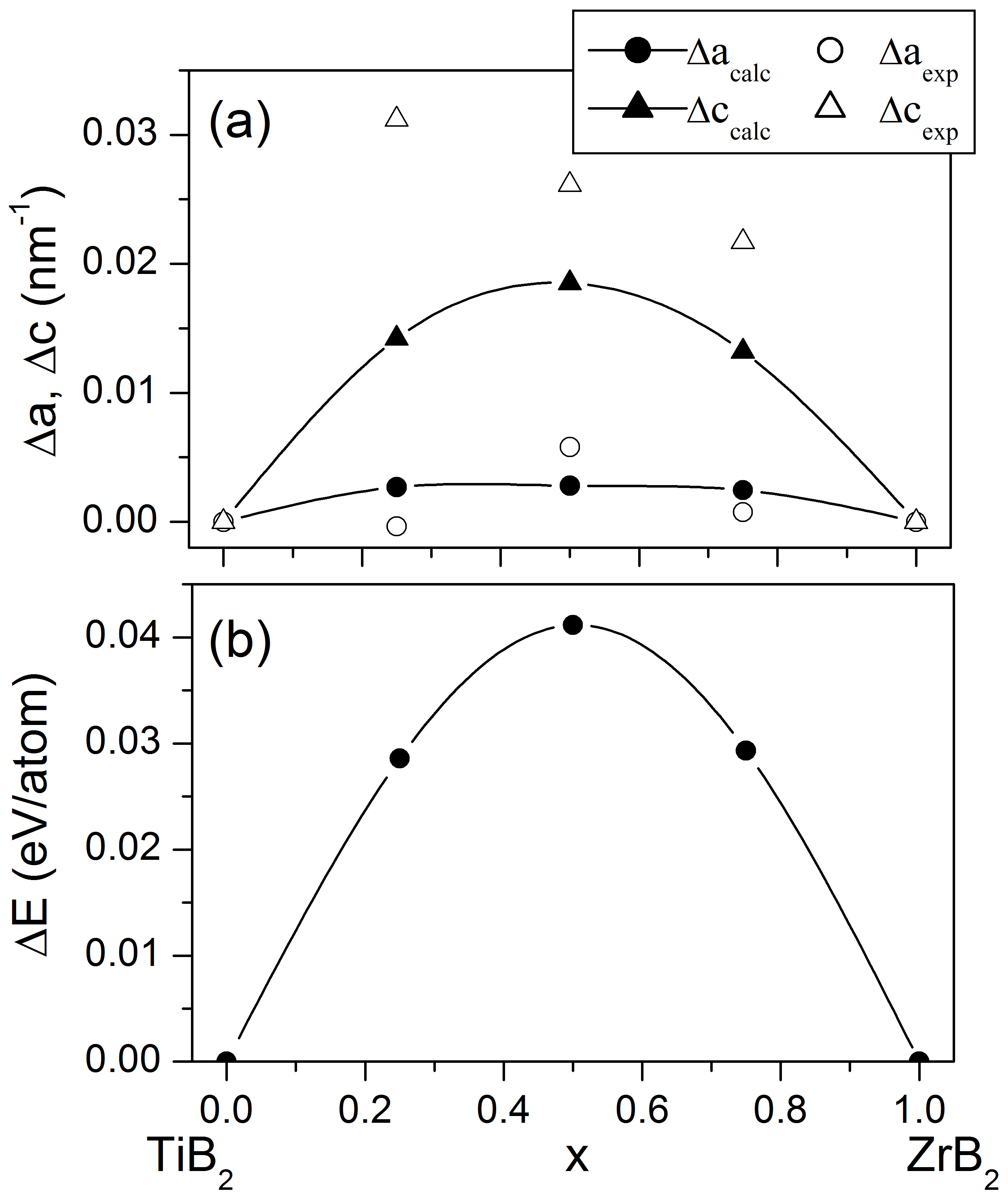}
  \caption{Deviation from linearity of the lattice parameters $\Delta a$ and
    $\Delta b$ (a) and mixing energy $\Delta E$ (b) for
    Ti$_{1-x}$Zr$_x$B$_2$.
    The calculated and experimental dependences are denoted by solid and
    dashed lines, respectively.
    Here, the lines are polynomial fits to the data points (to be considered
  as a guide to the eyes)~\cite{20}.}
  \label{fig:5}
\end{figure}

The random TiB$_2$-ZrB$_2$ solid solutions were investigated in
\cite{20} by using the first-principles calculation.
The results predict a positive deviation from Vegard's law for both the $a$ and
$c$ parameters (cf. figure~\ref{fig:5}a).
The positive curvature of the composition dependences of the lattice
parameters may point to the instability of Ti$_{1-x}$Zr$_x$B$_2$.
Indeed, the mixing energy for all compositions is positive (cf.
figure~\ref{fig:5}), which indicates that the solid solution is not stable and
will decompose into TiB$_2$ and ZrB$_2$ with the chemical driving force
$\Delta E(x)$.

\begin{figure}[htpb]
  \centering
  \includegraphics[width=0.5\linewidth]{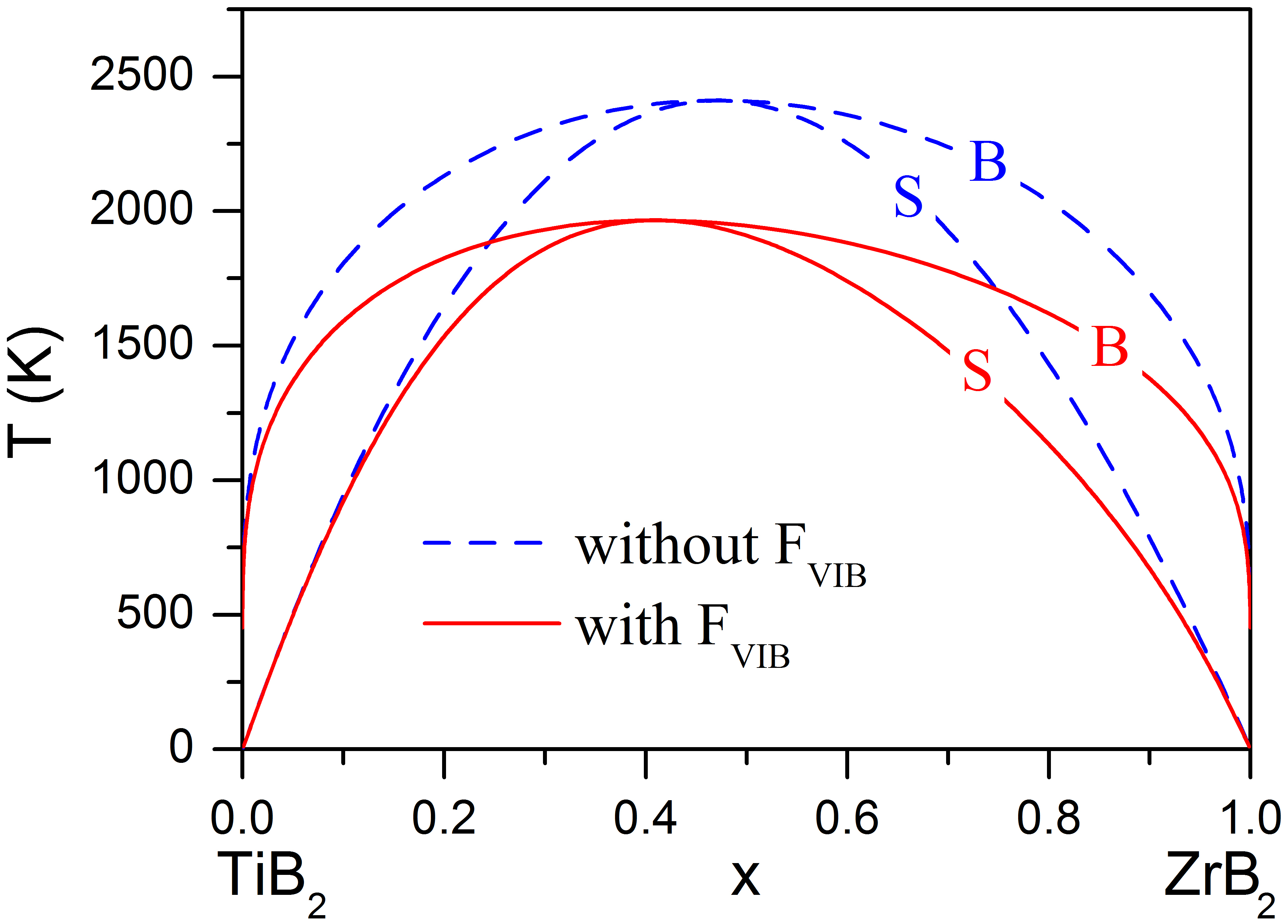}
  \caption{(Colour online) The calculated spinodal (S) and bimodal (B) curves for the
    Ti$_{1-x}$Zr$_x$B$_2$ alloys.
    Dashed blue curves are for calculations that did not include
    $F_\mathrm{Vib}$, and solid red curves are for calculations that
  did~\cite{20}.}
  \label{fig:6}
\end{figure}

\textcolor{black}{The phase diagrams of the TiB$_2$-ZrB$_2$ system, i.~e.,
  the concentration dependence of the decomposition temperature of the solid
  solutions according to the spinodal and binodal mechanisms, shown in
  figure~\ref{fig:6}, were computed with and without the contribution to the
  Gibbs free energy coming from vibration spectra
($F_\mathrm{Vib}$)~\cite{20}.}
For both the spinodal and bimodal curves calculated without $F_\mathrm{Vib}$,
\textcolor{black}{the maximum decomposition temperature (consolute temperature,
$T_\mathrm{C}$)} is 2418 K and the consolute composition ($x_\mathrm{C}$) is
0.47.
For the curves computed with contribution from $F_\mathrm{Vib}$,
$T_\mathrm{C} = 1973$ K and $x_\mathrm{C} = 0.4$. Therefore, it reduces the
consolute temperature and shifts both the spinodal and binodal curves
towards the TiB$_2$ (smaller-cation) side.

\begin{figure}[!t]
  \centering
  \includegraphics[width=0.39\linewidth]{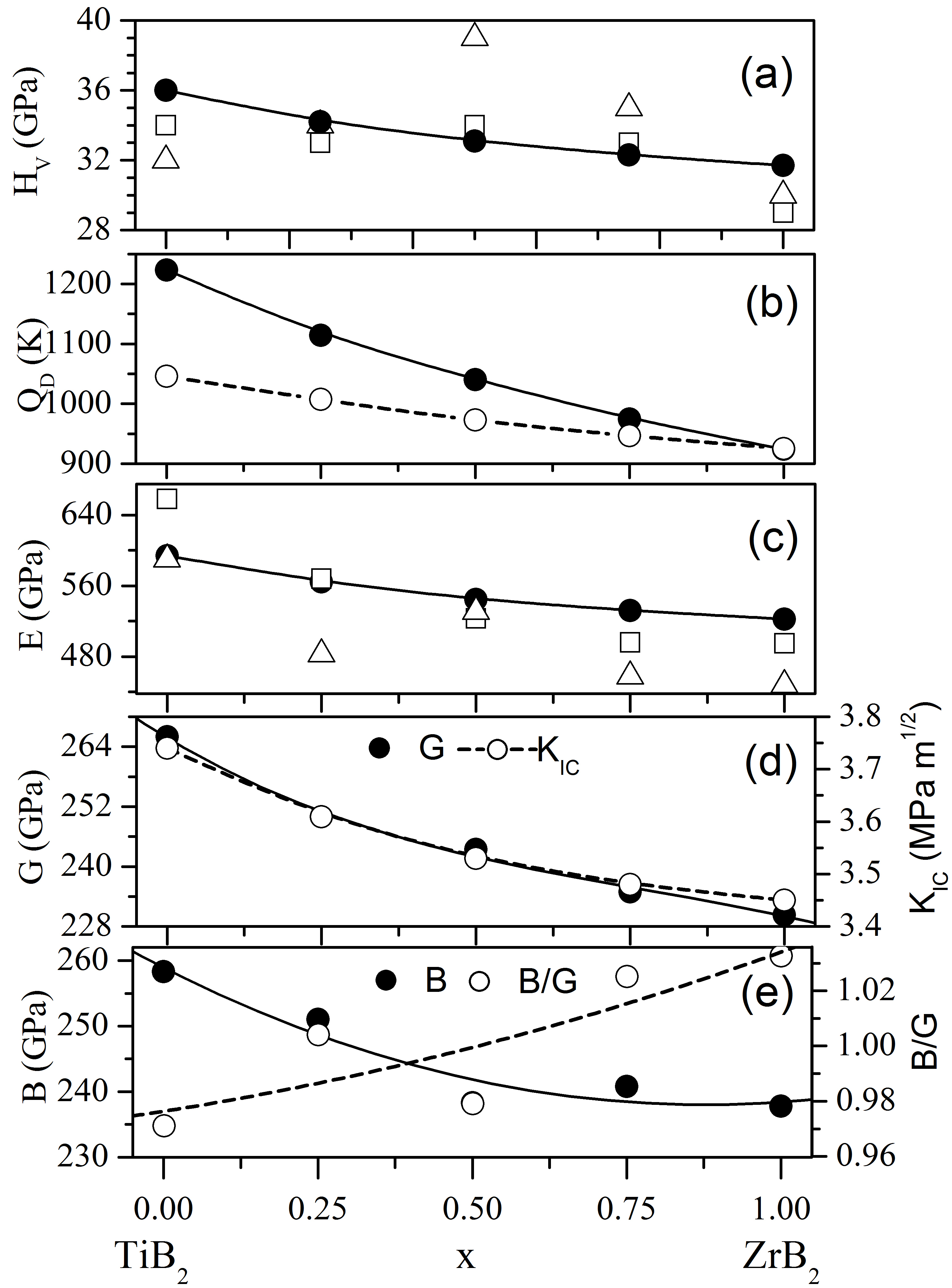}
  \caption{Vickers hardness ($H_\mathrm{V}$), Debye temperature
    ($Q_\mathrm{D}$) estimated using the elastic moduli (full circles) and
    the $C_V(T)$ dependences at $T$ = 300 K (open circles), Hill Young ($E$),
    shear ($G$), bulk ($B$) moduli, facture toughness ($K_\mathrm{IC}$) and
    $B/G$ ratio of Ti$_{1-x}$Zr$_x$B$_2$ as functions of composition.
    Open triangles and squares are the experimental data from
  Refs.~\cite{21,22}, respectively~\cite{20}.}
  \label{fig:7}
\end{figure}

The calculated Hill elastic moduli ($B$, $G$, $E$), Debye temperature
($Q_\mathrm{D}$), Vickers hardness ($H_\mathrm{V}$), $B/G$ ratio and fracture
toughness ($K_\mathrm{IC}$) of the TiB$_2$-ZrB$_2$ alloys are shown in
figure~\ref{fig:7} as functions of composition~\cite{20}.
A negative deviation of all the calculated characteristics from the mixing
rule is observed.
Poisson's ratio changes are in antiphase to the composition dependences of
the elastic moduli $G$ and $E$ (not shown here).
No strength enhancement was revealed: the hardness of the alloys does not
exceed $H_\mathrm{V}$ of TiB$_2$ since the $H_\mathrm{V}(x)$ dependence has
a negative deviation from linearity.
The fracture toughness for TiB$_2$ (3.74 MPa m$^{1/2}$) and ZrB$_2$ (3.45 MPa
m$^{1/2}$) is in a rather good agreement with the experimental values.
For TiB$_2$, the calculation slightly underestimates the experimental values
of $K_\mathrm{IC}$.
The $B/G$ ratio for TiB$_2$ and ZrB$_2$ was found to be 0.97 and 1.03,
respectively, and this ratio for the alloys did not exceed these values (cf.
figure~\ref{fig:7}e) indicating that the solid solutions should exhibit a brittle
behavior.

In \cite{23}, the calculated total energy, electronic and phonon
densities of states, elastic constants, hardness, shear and tensile ideal
strengths, thermodynamic values for the random Ti$_{1-x}$Nb$_x$B$_2$ alloys
were analyzed as functions of compositions.
The calculated mixing energy ($E_\mathrm{mix}$) of both Ti-Nb and Ti-Nb-B
solid solutions (alloys) shown in figure~\cite{8} indicate that it is
energetically favorable for Nb and Ti, as well as for NbB$_2$ and TiB$_2$ to mix and
form the stable alloys, in agreement with experiment~\cite{24}.
The minimum of the $E_\mathrm{mix}(x)$ dependence falls on $x\approx0.43$.
It follows from figure~\cite{8} that the main contribution to the stabilization
of the solid solutions comes from the direct Ti-Nb interactions, while the
Ti-B-Nb correlations shift the mixing energy minimum towards the compositions
enriched by titanium.

\begin{figure}[htpb]
  \centering
  \includegraphics[width=0.5\linewidth]{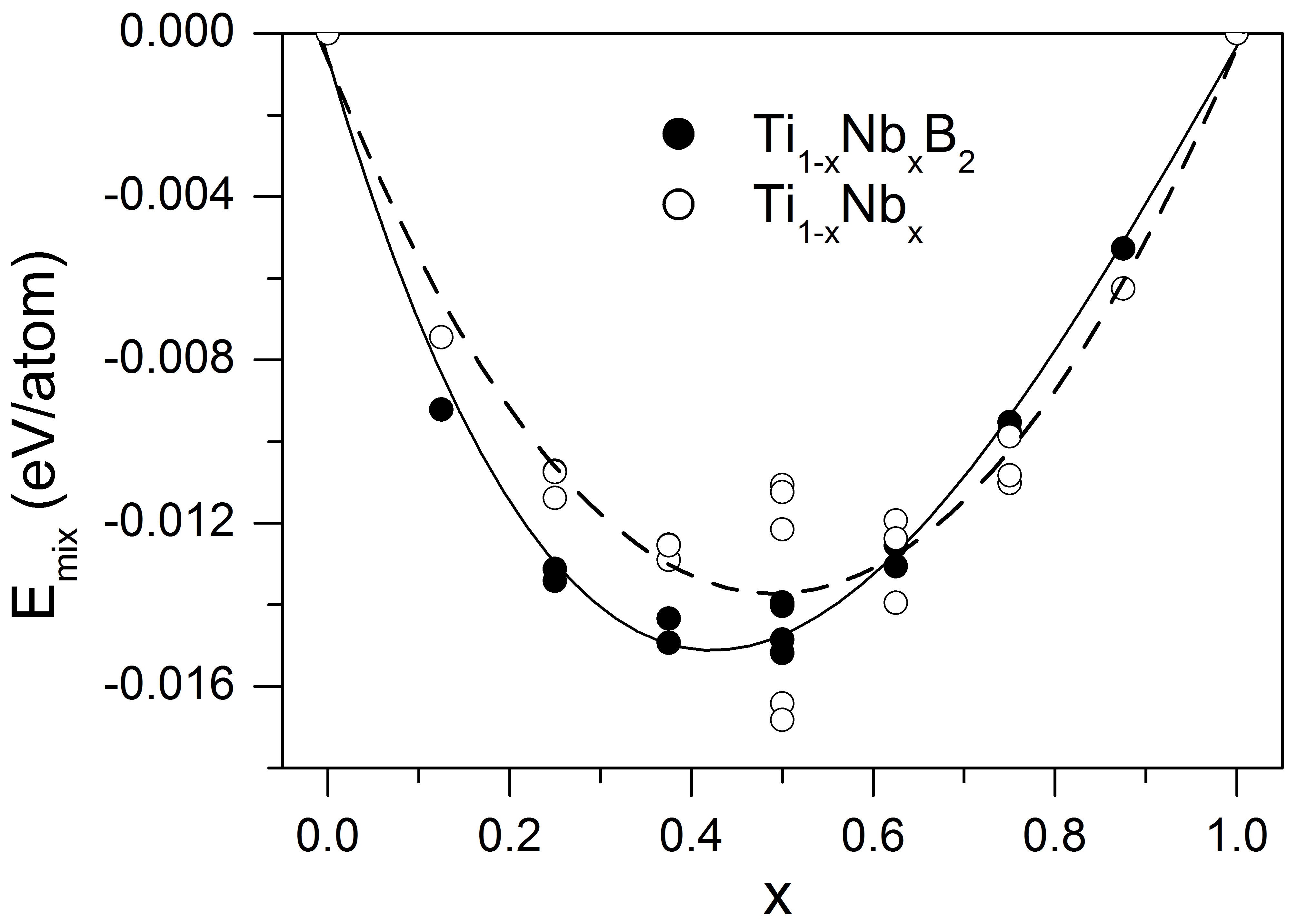}
  \caption{Energy of mixing versus composition for Ti$_{1-x}$Nb$_x$B$_2$
    (solid circles and solid line) and hypothetical Ti$_{1-x}$Nb$_x$ (open
    circle and dashed line).
    The lines are the resulting polynomial fit to the calculated
  points~\cite{23}.}
  \label{fig:8}
\end{figure}

\begin{figure}[htpb]
  \centering
  \includegraphics[width=0.4\linewidth]{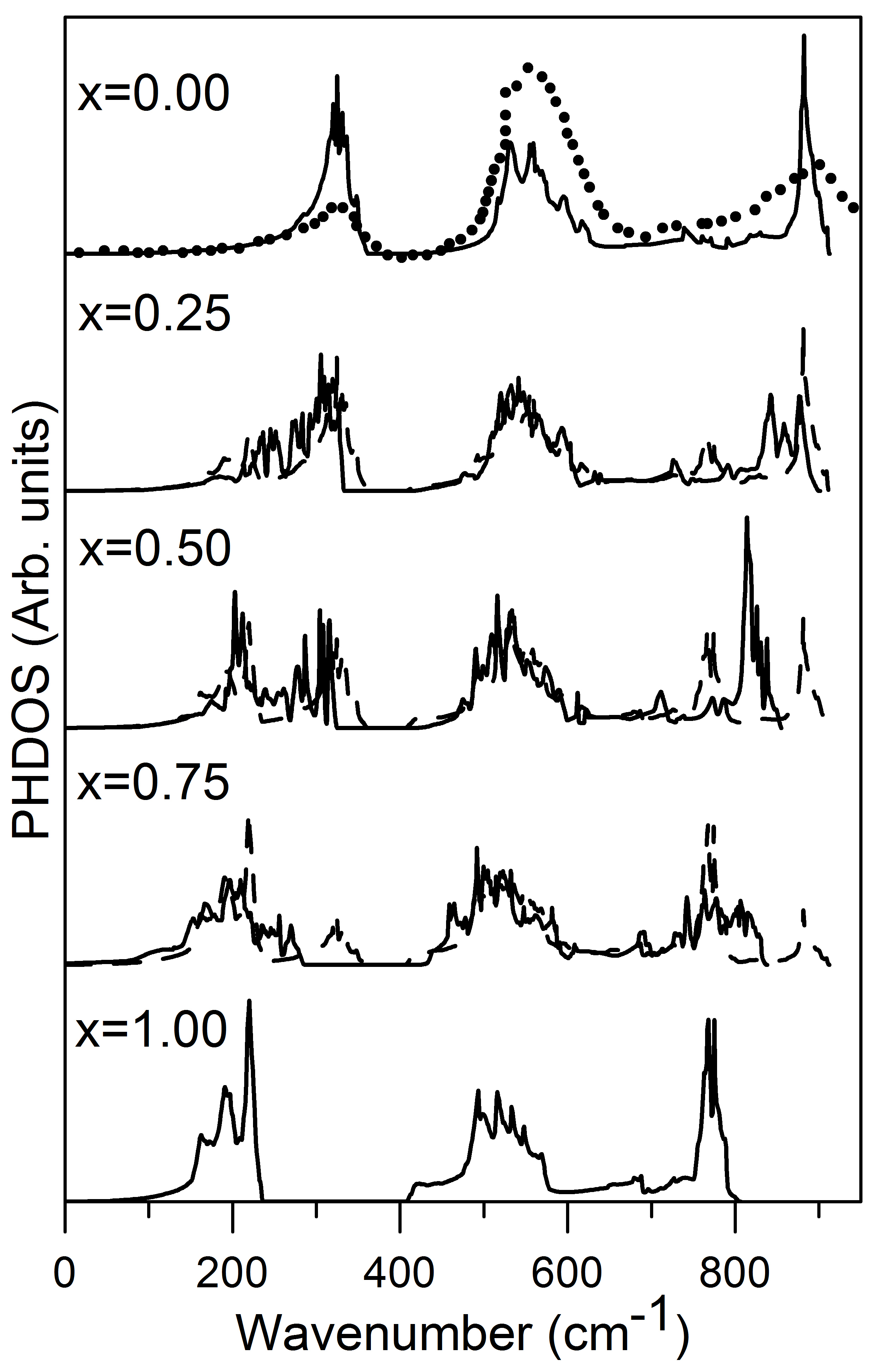}
  \caption{The calculated phonon density of states (PHDOS) for
    Ti$_{1-x}$Nb$_x$B$_2$ (solid line) and the composition weighted average
    PHDOSs of TiB$_2$ and NbB$_2$,
    $g_\mathrm{av} = (1-x) g_{\mathrm{TiB}_2} + x g_{\mathrm{NbB}_2}$
    (dashed line)~\cite{23}.
    For comparison, the experimental PHDOS measured  using the technique of
    inelastic neutron scattering for TiB$_2$ is also presented (dotted
  line) \cite{25}.}
  \label{fig:9}
\end{figure}

To clarify the role of the vibration spectra in the stabilization of the
solid solutions, the phonon densities of states (PHDOS, $g(x)$) of
Ti$_{1-x}$Nb$_x$B$_2$ and the composition weighted average PHDOSs of TiB$_2$
and NbB$_2$, 
\textcolor{black}{$g_\mathrm{av} = (1-x) g_{\mathrm{TiB}_2} +
x g_{\mathrm{NbB}_2}$}
were calculated and the results are shown in figure~\ref{fig:9}~\cite{23}.
For TiB$_2$, both the experimental and the calculated PHDOSs agree rather well.
The difference between PHDOSs for Ti$_{1-x}$Nb$_x$B$_2$
($g_\mathrm{alloy}(x)$, solid line) and the composition weighted average
PHDOSs of TiB$_2$ and NbB$_2$ (\textcolor{black}{$g_\mathrm{av}(x)$,} dashed
line) for each composition of the alloys takes place:
\textcolor{black}{$g_\mathrm{av}(x)$} shifts towards larger $x$.
However, the larger the PHDOS at large frequencies is, the larger (more
positive) the contribution to the vibration free energy becomes.
It follows that lattice vibrations will promote the stabilization of
Ti$_{1-x}$Nb$_x$B$_2$ at finite temperatures.

\begin{figure}[!t]
  \centering
  \includegraphics[width=0.5\linewidth]{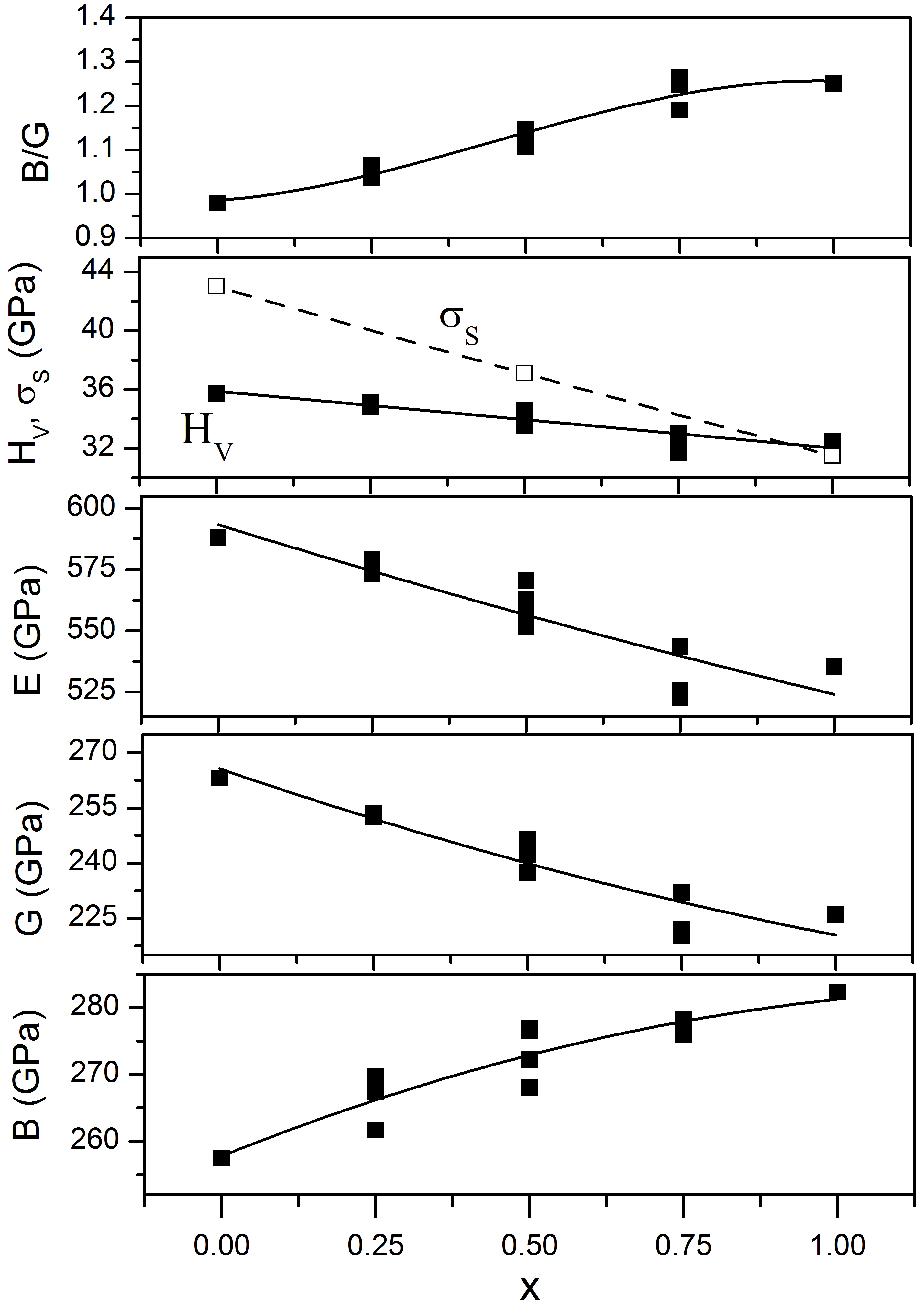}
  \caption{The calculated bulk modulus ($B$), elastic constants ($G$, $E$),
    Vickers hardness ($H_\mathrm{V}$), ideal shear strength
    \textcolor{black}{($\sigma_s$)}, and $B/G$ ratio for Ti$_{1-x}$Nb$_x$B$_2$
    alloys as functions of concentration $x$ (squares).
    The solid lines are the resulting cubic interpolation of the calculated
  points \cite{23}.}
  \label{fig:10}
\end{figure}

The calculated elastic moduli, bulk-to-shear modulus ratio ($B/G$) and
Vickers hardness for Ti$_{1-x}$Nb$_x$B$_2$ are shown in figure~\ref{fig:10} as
functions of composition.
One can see that the calculated moduli for TiB$_2$ and NbB$_2$ agree well
with those obtained in other experimental and theoretical studies.
For the alloys, we see that the calculated bulk modulus increases, and the
values of $G$, $E$ and $H_\mathrm{V}$ decrease gradually with an increasing $x$.
An extreme behavior of these values with composition was not observed: the
solid solutions have always got lower values of the elastic constants and moduli
than the composition weighted average values of TiB$_2$ and NbB$_2$,
suggesting that the elastic properties are restricted by the individual
components.
To evaluate the alloying effect on the ductility, the ratio, $B/G$, as a
function of concentration $x$ is plotted in figure~\ref{fig:10}.
The higher or lower the $B/G$ ratio is, the more ductile or brittle the
material is, respectively.
The critical value that separates ductile and brittle materials is
approximately 1.75. Figure~\ref{fig:10} shows that the $B/G$ ratio increases
with $x$, although it does not reach the critical value.
Based on the trends for the $B/G$ ratio, one can conclude that the ductility
of the alloys increases gradually as the Nb content increases.

In \cite{26}, the stability, electronic structures and mechanical
properties of Mo$_2$FeB$_2$ and Mo$_2$NiB$_2$ ternary borides were
investigated.
Mo$_2$FeB$_2$ was found to display high hardness and strength as well as
pretty good corrosion-resistance, which attracts much attention compared to
other borides~\cite{27}.
Mo$_2$NiB$_2$ was found to have two different structures with orthorhombic
and tetragonal lattices (O-Mo$_2$NiB$_2$ and T-Mo$_2$NiB$_2$, respectively).
Compared to Mo$_2$FeB$_2$, the Mo$_2$NiB$_2$ based cermets are supposed to
show better high-temperature properties and corrosion resistance considering
the existence of Ni.
In this context, Mo$_2$NiB$_2$ is hypothetically more appropriate for the
applications under the conditions of high temperature.

\begin{figure}[!t]
  \centering
  \includegraphics[width=0.66\linewidth]{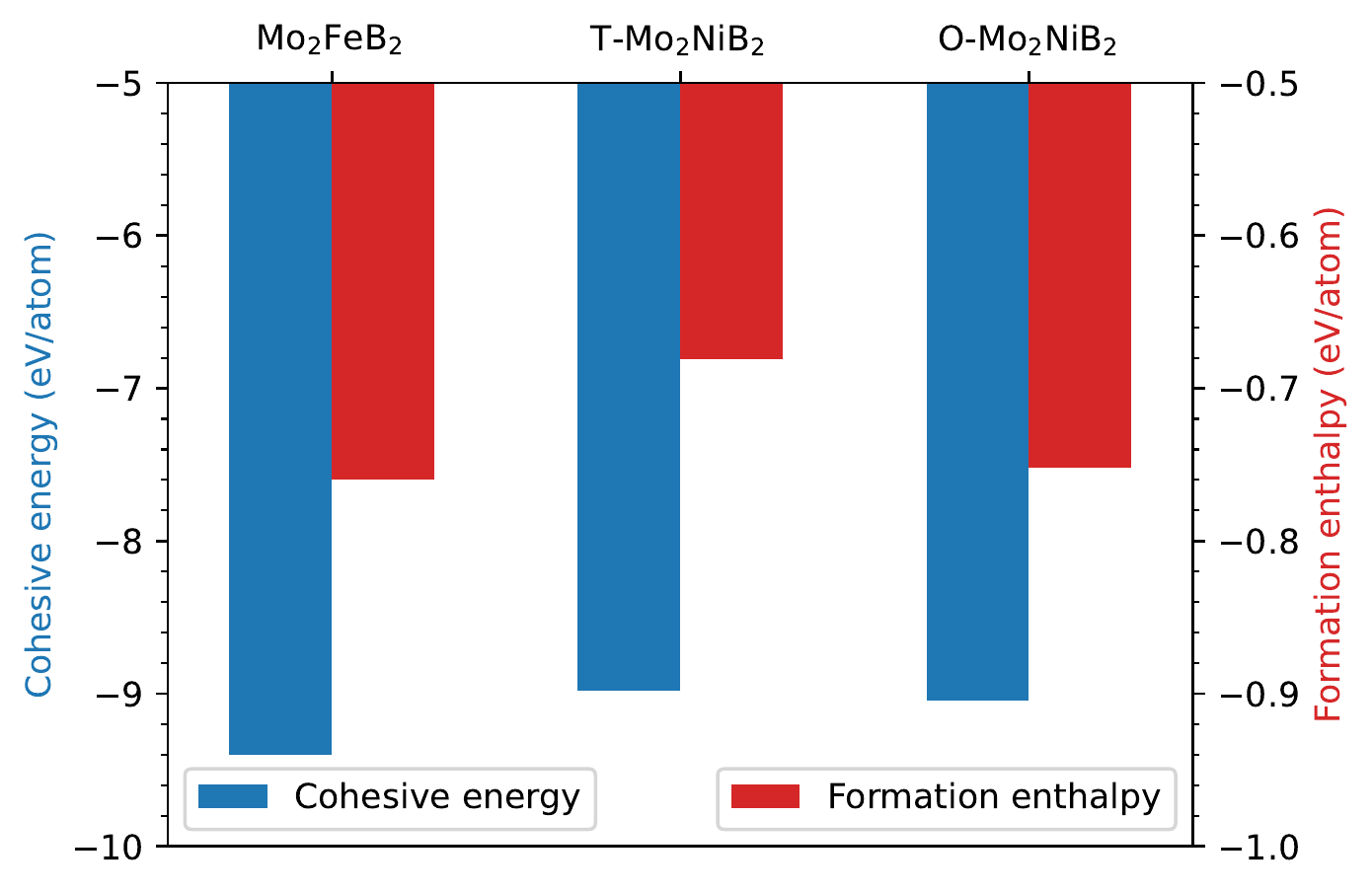}
  \caption{(Colour online) Cohesive energy and formation enthalpy of three ternary borides:
  Mo$_2$FeB$_2$, T-Mo$_2$NiB$_2$, O-Mo$_2$NiB$_2$ \cite{26}.}
  \label{fig:11}
\end{figure}

Shown in figure~\ref{fig:11} are the calculated cohesive energy and formation
enthalpy of three ternary borides.
It can be seen that the cohesive energy and formation enthalpy are negative
indicating that all the three borides are thermodynamically stable.

\begin{figure}[!t]
  \centering
  \includegraphics[width=0.66\linewidth]{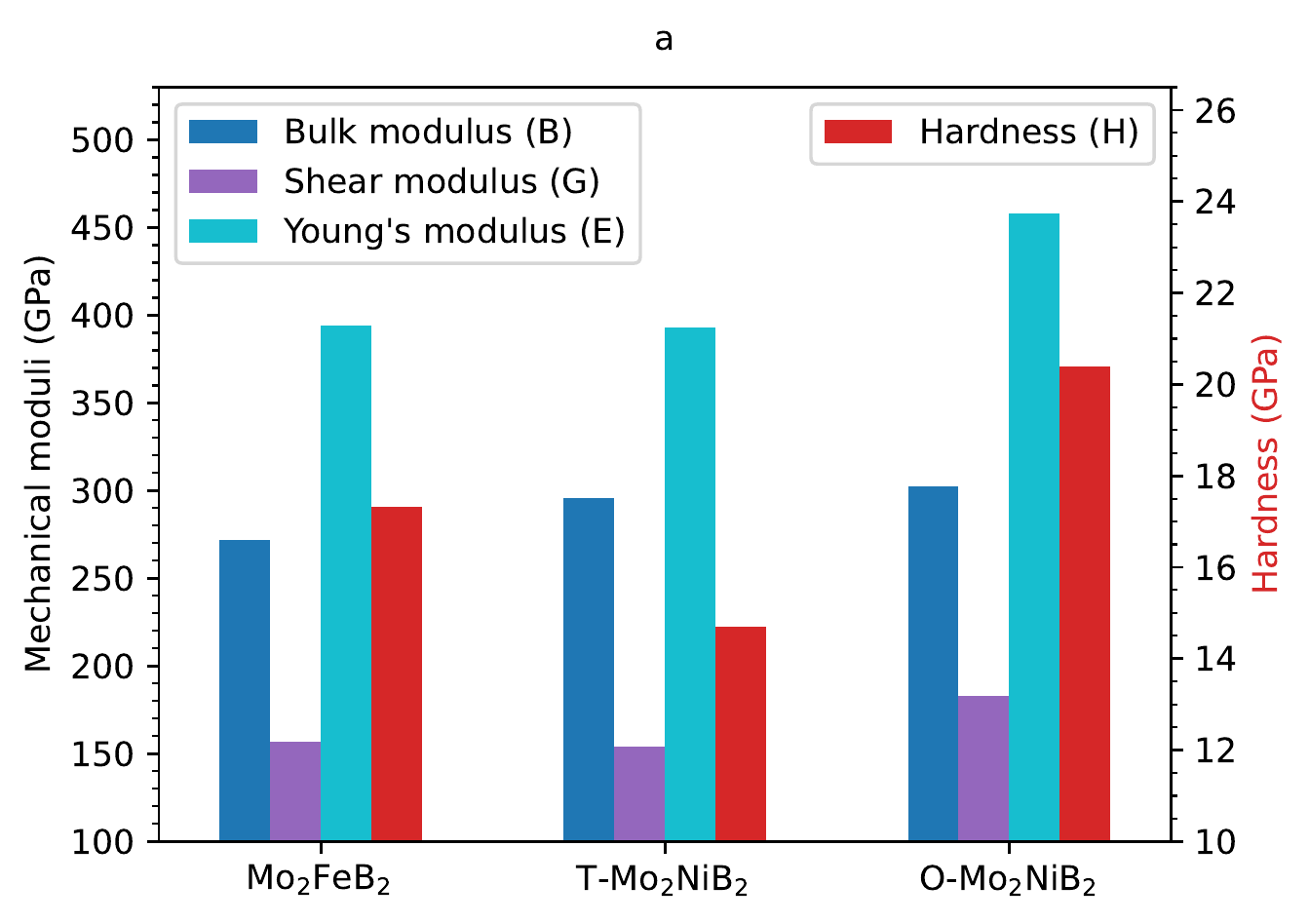} \\
  \includegraphics[width=0.66\linewidth]{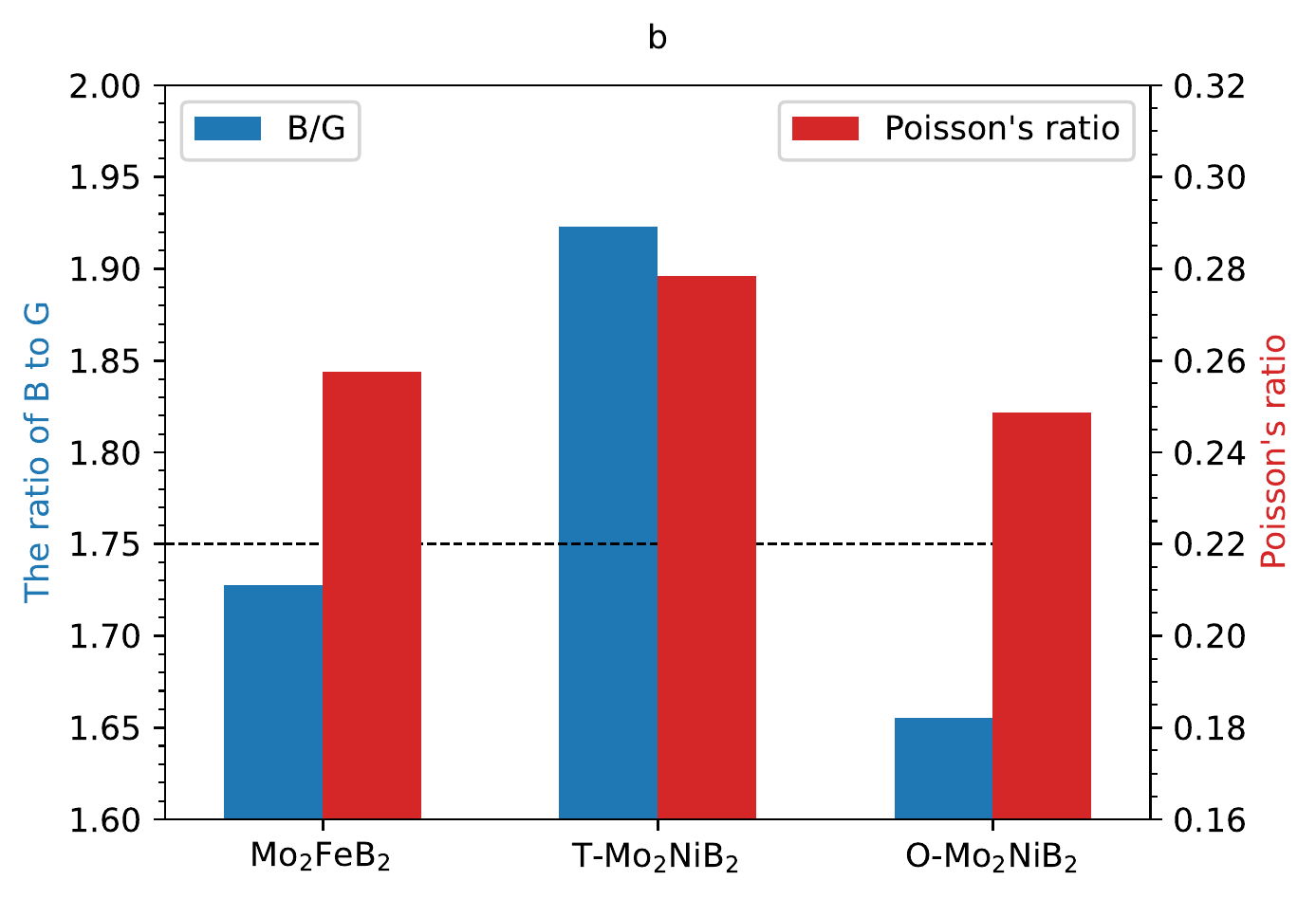} \\
  \caption{(Colour online) Mechanical properties of three ternary borides:
    (a) Moduli and \textcolor{black}{Vickers} hardness,
  (b) the ratio of $B/G$ and Poisson's ratio~\cite{26}.}
  \label{fig:12}
\end{figure}

It can be found from figure~\ref{fig:12}a that the hardness shows a good
consistence to the shear modulus of the three ternary borides.
Young's modulus is defined as the ratio of stress to strain, which can
reflect the stiffness of the solid.
In this case, it can be concluded that O-Mo$_2$NiB$_2$ shows the highest
stiffness with Mo$_2$FeB$_2$ and T-Mo$_2$NiB$_2$ followed by sequence.
Moreover, Young's modulus can respond to the covalent bonding character of the
material.
Higher Young's modulus indicates a stronger covalent bonding.
Thus, O-Mo$_2$NiB$_2$ can be considered to have the strongest covalent bonding
while Mo$_2$FeB$_2$ and T-Mo$_2$NiB$_2$ show a relatively weaker covalent
bonding character.
Furthermore, it can be found that \textcolor{black}{the Vickers hardness to
Young modulus ($H/E$) ratio} of O-Mo$_2$NiB$_2$ is slightly higher than that
of Mo$_2$FeB$_2$ and T-Mo$_2$NiB$_2$.
Figure~\ref{fig:12}b shows the $B/G$ ratio and Poisson's ratio of the three
 ternary borides studied.
Generally, $B/G$ ratio can be used to indicate the ductility of the material.
It can be found that O-Mo$_2$NiB$_2$ exhibits the smallest $B/G$ ratio,
implying the weakest ductility.
By comparison, T-Mo$_2$NiB$_2$ shows the best ductility with the $B/G$ ratio
of 1.923.
Usually, the material possessing such high $B/G$ can be classified as a ductile
compound.
As mentioned above, O-Mo$_2$NiB$_2$ has a strong tendency to covalent bonding,
which may be the main factor for its brittleness.
On the contrary  T-Mo$_2$NiB$_2$ shows a stronger tendency to metallic character,
which is attributed to the typical tetragonal crystal structure.

\textcolor{black}{Thus, the main results of the investigations
  of Mo$_2$FeB$_2$ and Mo$_2$NiB$_2$ ternary borides can be summarized as
follows:}
(1) Mo$_2$FeB$_2$, T-Mo$_2$NiB$_2$ and O-Mo$_2$NiB$_2$ are all
thermodynamically and mechanically stable, and O-Mo$_2$NiB$_2$ shows a better
stability than T-Mo$_2$NiB$_2$.
(2) O-Mo$_2$NiB$_2$ shows the highest Young's modulus and hardness such as 457.3
and 20.4 GPa, respectively; T-Mo$_2$NiB$_2$ has a comparable hardness and
modulus to Mo$_2$FeB$_2$ while the ductility is obviously higher.
(3) O-Mo$_2$NiB$_2$ shows the best isotropy among the three studied borides
though it shows the strongest covalent bonding character.

\section{Methods for deposition of ternary transition metal boride \\films/coatings}

\begin{figure}[!b]
  \centering
  \includegraphics[width=0.4\linewidth]{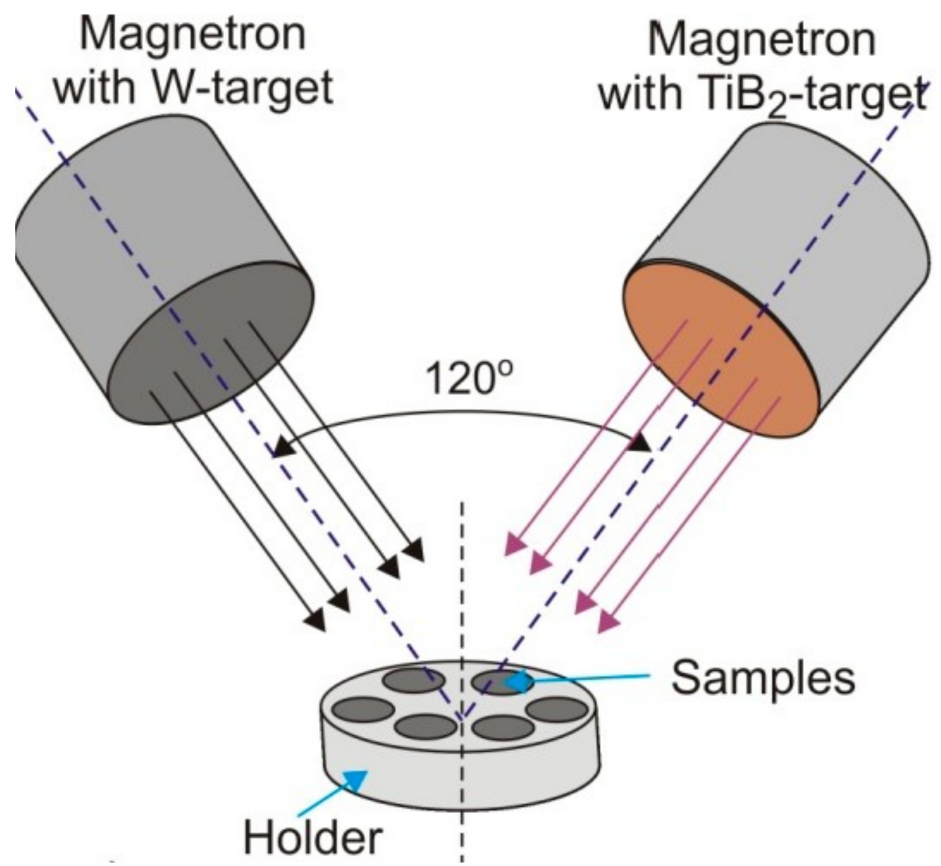}
  \caption{(Colour online) Scheme of the unit equipped with two magnetrons~\cite{28}.}
\label{fig:13}
\end{figure}

The main properties of condensates depend not only on their structure and
chemical bonds, but also, to a great extent indirectly, on the deposition
process itself.
The films/coatings of transition metal ternary borides were deposited
using various techniques.
The most commonly used technique for depositing ternary transition boride
films is magnetron sputtering in which either one target prepared from
mixture of respective elemental or binary boride powders~\cite{31,32,33}, or
co-deposition of two targets each prepared only by one of the binary borides
which are constituents of the deposited film~\cite{28,29,30,34,36,37,39} are
used.
In the latter case, the units equipped with two magnetrons are suitable for the
``dual'' deposition process (figure~\ref{fig:13}).

In \cite{35}, the Fe-Cr-B coatings were deposited by detonation-gun
thermal spraying method of powders of Fe-Cr-B alloy.
The acetylene (C$_2$H$_2$) and oxygen (O$_2$) were used as  fuel gases.
In \cite{38}, the Ni-Cr-B coatings were prepared by this method through
spraying NiCrB cored wires prepared by filling NiCr alloy tubes with the
mixture of NiB, Cr and Ni powders.

\begin{figure}[!b]
  \centering
  \includegraphics[width=0.6\linewidth]{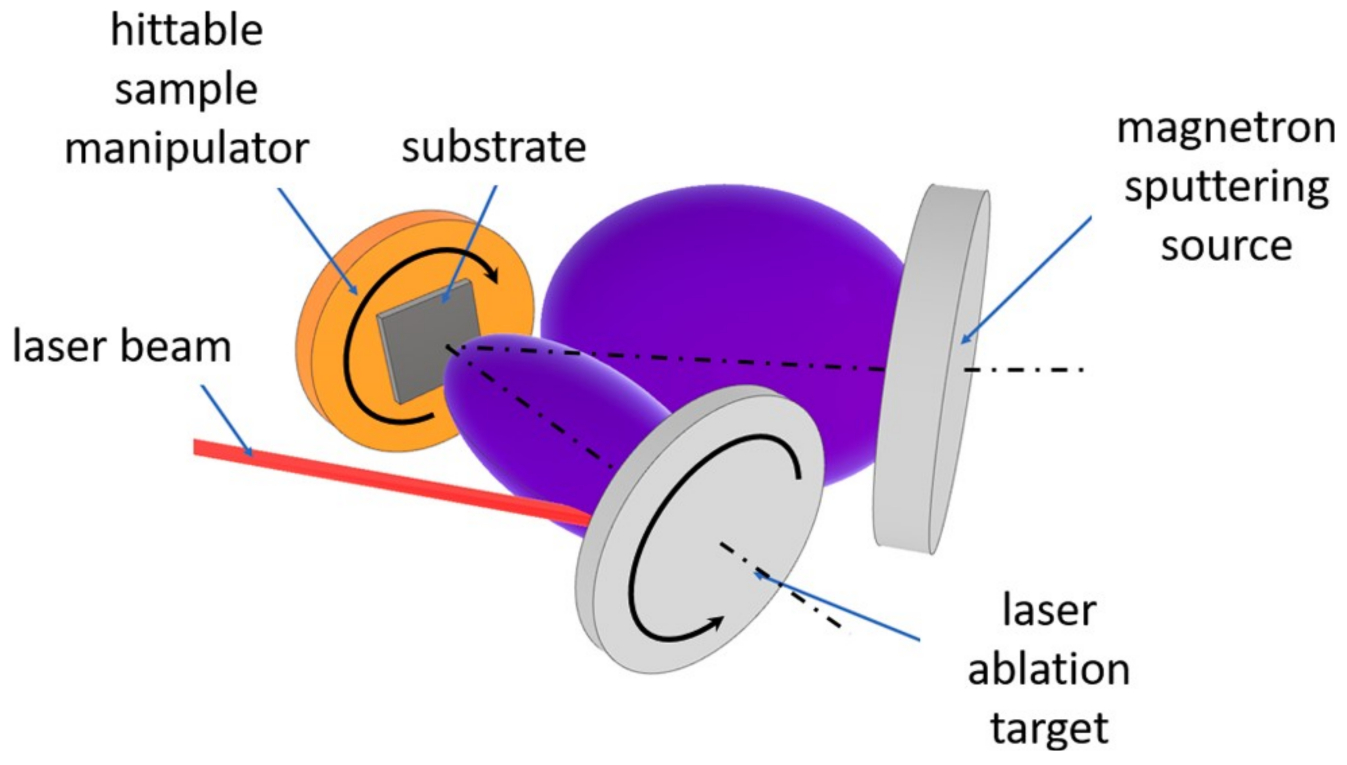}
  \caption{(Colour online) Scheme of magnetron sputtering combined with pulsed laser
  deposition~\cite{29}.}
  \label{fig:14}
\end{figure}

In \cite{29}, the films of W-Zr-B were deposited by hybrid RF magnetron
and pulsed laser method.
The sputtering gas was argon.
In this case, the target made of W$_2$B$_5$ was sputtered by magnetron whereas
the target made of ZrB$_2$ was sputtered by laser (figure~\ref{fig:14}).
One more technique used for depositing the ternary transition metal boride films
is the arc spraying method.

\section{Structure}

The structure and elemental and phase composition of ternary boride films
were studied by X-ray diffraction (XRD), time of flight elastic recoil
detection analysis (TOF-ERDA), scanning electron microscopy (SEM),
energy-dispersive spectroscopy (EDS), wavelength-dispersive spectroscopy
(WDS), electron-probe microanalysis (EPMA), transmission electron microscopy
(TEM), selective area electron diffraction (SAED), X-ray photoelectron
spectroscopy (XPS), elastic recoil detection (ERD) and Rutherford
backscattering (RBS), electron energy-loss spectroscopy (EELS).

\begin{figure}[htpb]
  \centering
  \includegraphics[width=0.45\linewidth]{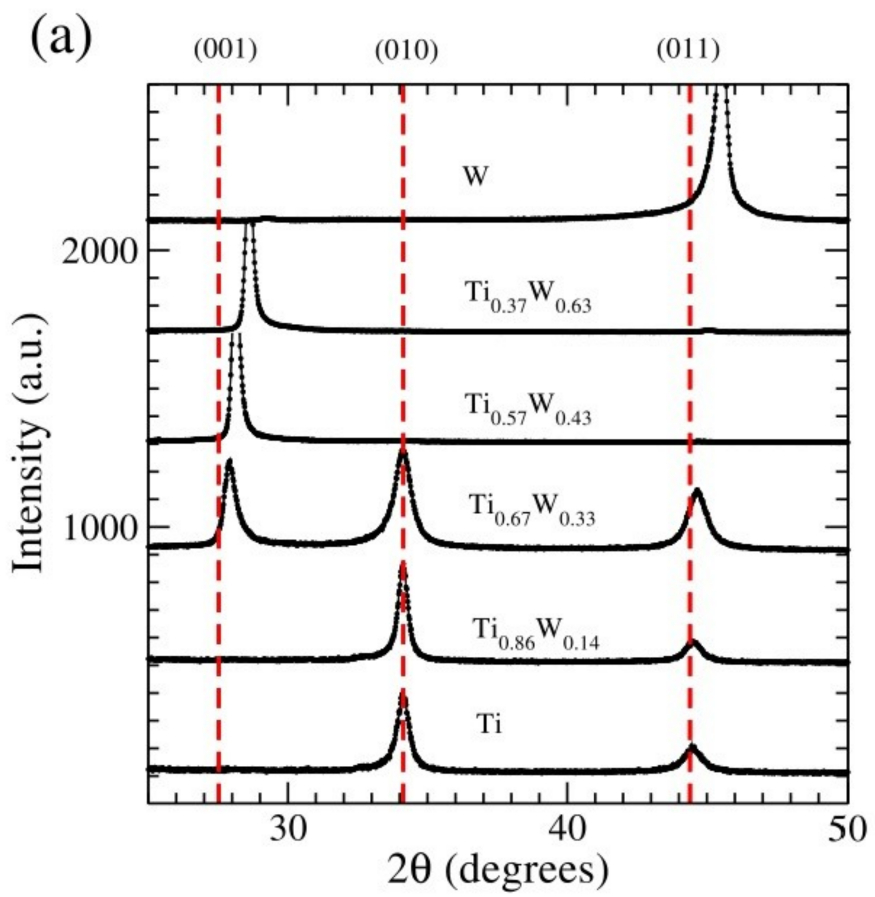}
  \caption{(Colour online) Diffraction patterns (grazing incidence) of Ti$_x$W$_{1-x}$B$_2$
  thin films~\cite{30}.}
  \label{fig:15}
\end{figure}

The films in Ti-W-B system with different C$_\mathrm{Ti}$-to-C$_\mathrm{W}$
ratio were studied in~\cite{28,30,31,32}.
In~\cite{30}, the effect of tungsten addition on the microstructure and
properties of the Ti$_x$W$_{1-x}$B$_2$ thin films was studied.
Shown in figure~\ref{fig:15} are typical XRD patterns of the Ti$_x$W$_{1-x}$B$_2$
thin films deposited on Si substrates~\cite{30}.
XRD results show that all films are single phased and have crystallized ---
even at high tungsten content --- in the AlB$_2$ structure type.
With increasing tungsten content, both the (001) and the (011) reflections
shift toward higher diffraction angles, whereas the (010) reflection remains
at the position obtained for TiB$_2$.
This indicates that replacing Ti by W results in the formation of (Ti,W)B$_2$
solid solution.
Note that the content of Ti in the films studied was 12.8$\div$28.3 at. \%.
The shift of base reflections toward higher or lower diffraction angles was
observed also in W-Zr-B~\cite{29}, W-Ti-B~\cite{31,32} and Zr-Ta-B~\cite{37}
films.
The shift of XRD reflections toward higher or lower diffraction angles
depends on  whether the adding atoms have larger or smaller radius 
compared to the base atoms.
In these cases, the introduced atoms replace the base atoms and cause an
increase or decrease of the unit cell (i. e., lattice parameter) due to
compressive or tensile stress, and as a result we observe a shift of diffraction peaks.

\begin{figure}[htpb]
  \centering
  \includegraphics[width=0.25\linewidth]{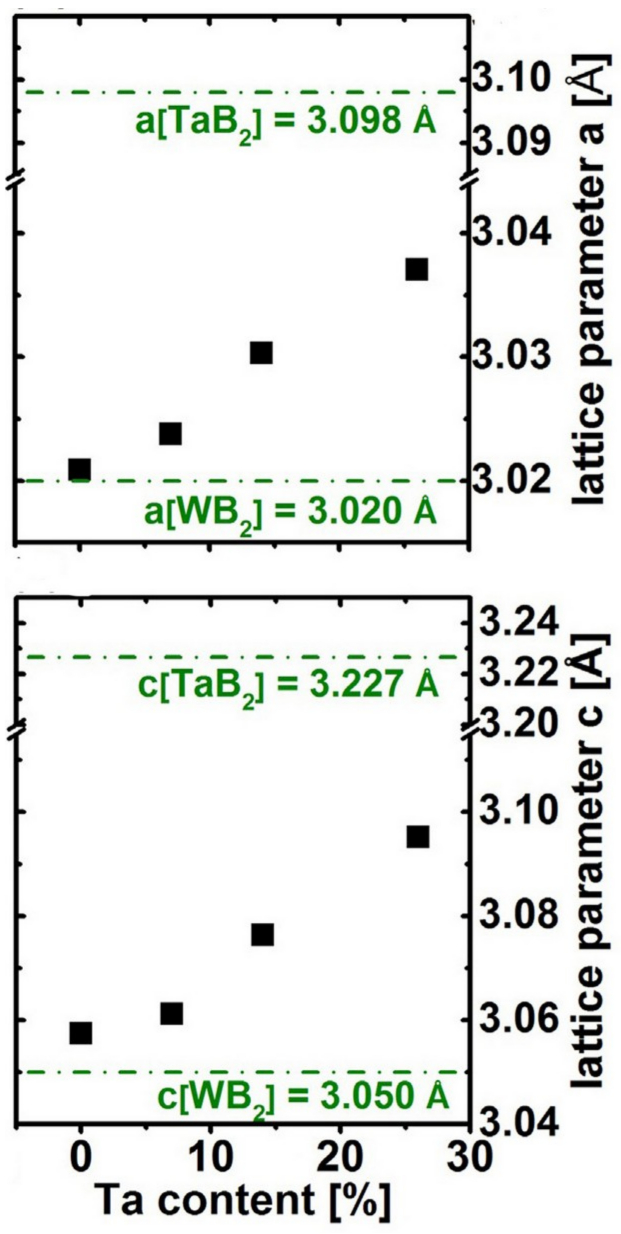}
  \caption{(Colour online) Lattice parameters ($a$ and $c$) of the single-phased coatings
    (annealed at $T_\mathrm{a}$ = 800\textcelsius{} for 16 h in vacuum to ensure
  stress free states) vs. Ta content~\cite{33}.}
  \label{fig:16}
\end{figure}

The Ti-W-B films with different content of Ti (0.6$\div$12.2 at. \%) were
also studied in~\cite{31}. It was revealed that at low Ti content 
(0.6 at. \% $\leq$ Ti $\leq$ 4.6 at. \%) there is formed a coherently bound nanocomposite
$\beta$-(W,Ti)B/(Ti,W)B$_2$.
At a content of 10.2 at. \% Ti, the structure of the coating is ordered on
the basis of (Ti,W)B$_2$ phase.
XRD pattern  also showed that with an increasing Ti content, the (101)
(Ti,W)B$_2$ diffraction peak became relatively more intense.
This indicated the presence of a preferred orientation of the grains in the
films.
The average crystallite size of the $\beta$-(W,Ti)B phase decreased in the
interval C$_\mathrm{Ti}$/C$_\mathrm{W}$ < 0.15.
The observed change in the size of the crystallites is explained by the
formation of the $\beta$-(W,Ti)B solid solution.
When the content of titanium in the film is high, the size of the
$\beta$-(W,Ti)B crystallites increases proportional to the ratio
C$_\mathrm{Ti}$/C$_\mathrm{W}$.
A linear increase in the lattice parameter with an increase in concentration of Ta
in W-Ta-B films (figure~\ref{fig:16}) was observed in~\cite{33} with an excellent
agreement to Vegard's relationship for substitutional solid
solutions~\cite{40}.
The formation of solid solution single-phase AlB$_2$-type TM boride films through
partial replacing of the matrix TM atoms by the added TM atoms was also
observed in W-Zr-B~\cite{29}, W-Ti-B [32], W-Ta-B~\cite{33,34},
Fe-Cr-B~\cite{35}, Ti-Ta-B~\cite{36}, Zr-Ta-B~\cite{37}, and
Ni-Cr-B~\cite{38} ternary borides.

\begin{figure}[htpb]
  \centering
  \includegraphics[width=0.99\linewidth]{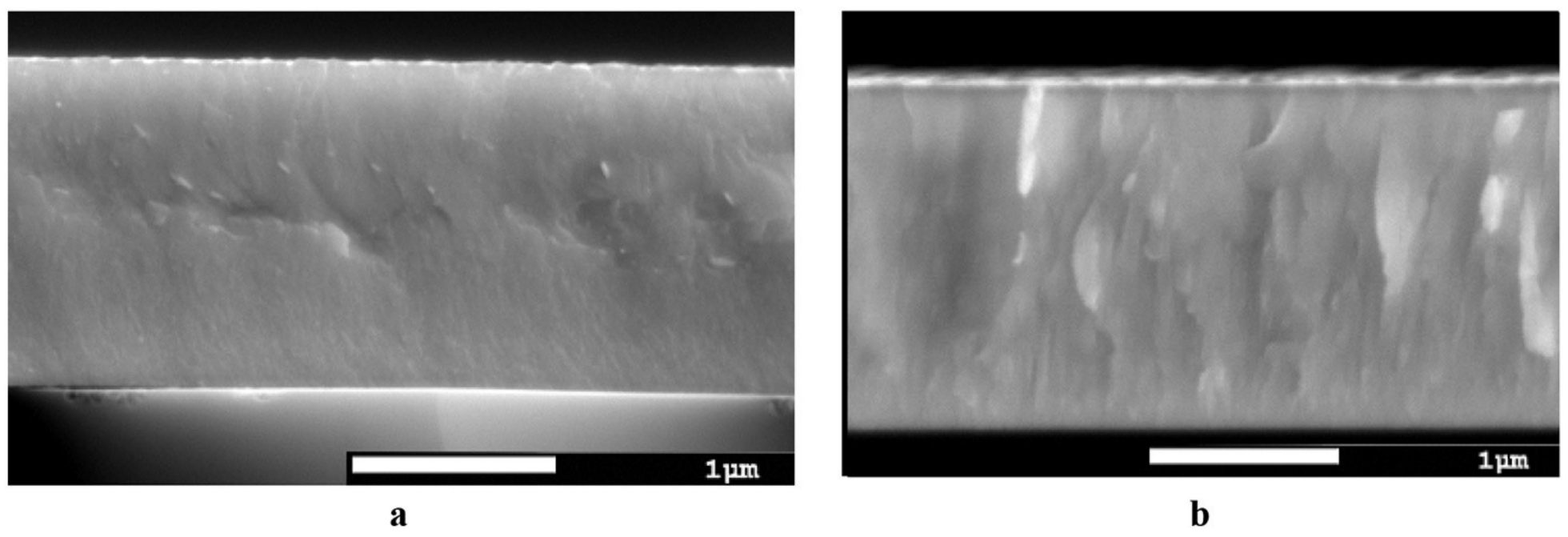}
  \caption{Cross-section SEM images of W-Ti-B films with content Ti, at. \%:
  a --- 1.9, b --- 10.2~\cite{31}.}
  \label{fig:17}
\end{figure}

\begin{figure}[htpb]
  \centering
  \includegraphics[width=0.99\linewidth]{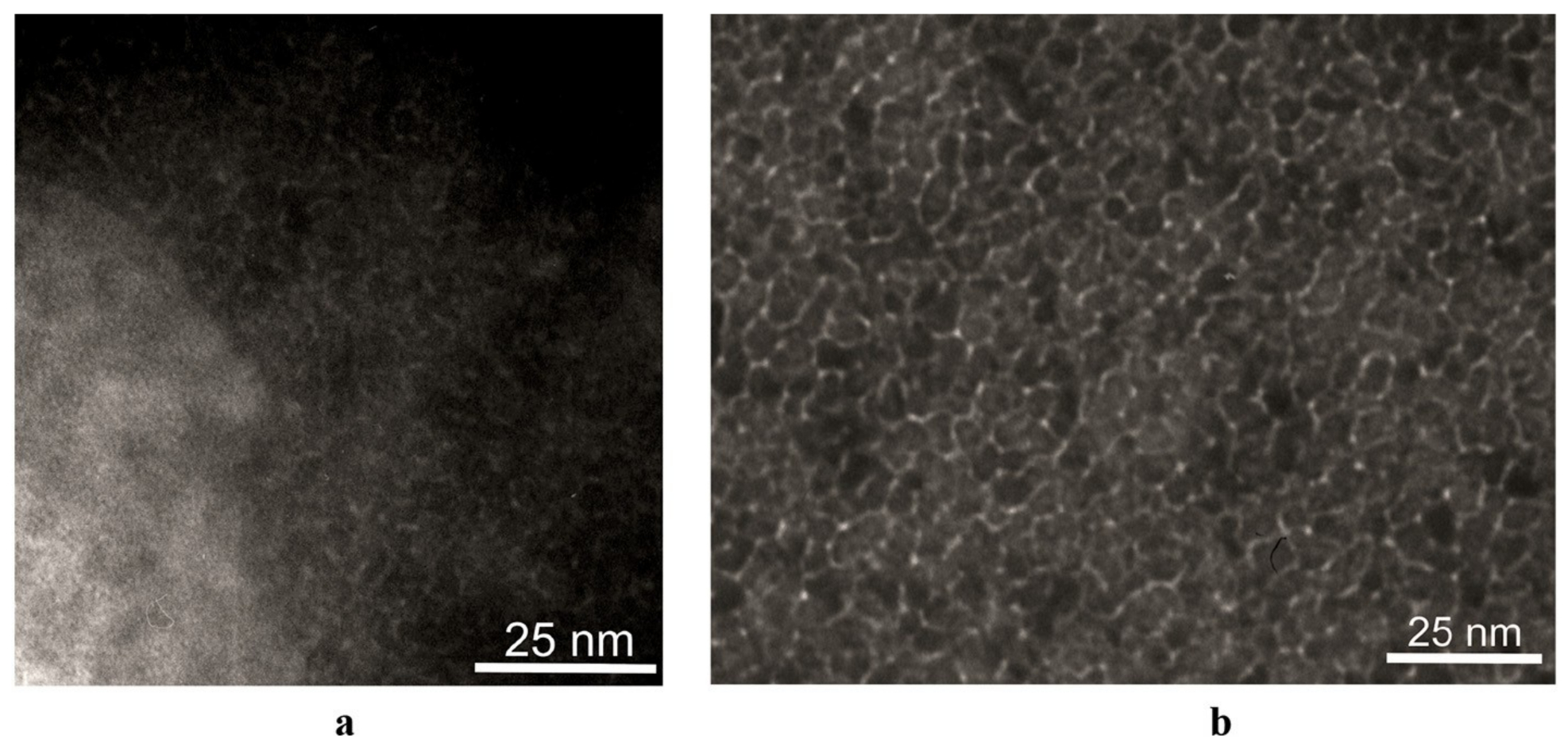}
  \caption{HR TEM images of W-Ti-B film with content Ti, at. \%:
  a --- 1.9, b --- 10.2~\cite{31}.}
  \label{fig:18}
\end{figure}

The structure of Ti-W-B films was also studied using SEM and TEM
methods~\cite{28,30,31,32}.
Formation of an oriented grain structure in those films was observed.
For instance, in \cite{31} high-resolution SEM \textcolor{black}{images}
of the cross-sectional area of Ti-W-B films showed that for a small content
of titanium atoms in films, an unoriented structure is mainly formed
(figure~\ref{fig:17}a).
At a large content of Ti (C$_\mathrm{Ti}$ $\approx$ 10.2 at. \%), a columnar
growth of crystallites occurred in the films (figure~\ref{fig:17}b).
The high resolution TEM images of those films are shown in figure~\ref{fig:18}.
As can be seen from the figure, in this case an ordered structure is formed
(the average size of the ordering regions is about 10 nm).
Formation of an oriented grain structure was also observed in W-Zr-B~\cite{29},
W-Ta-B~\cite{33}, Zr-Ta-B~\cite{37} and La-Zr-B~\cite{39} films.

The XPS method was used to study the chemical bonds in ternary TM borides.
In \cite{29}, the state of W, Zr and B bonds in the W-Zr-B films was
investigated.
It was revealed that only W-B and Zr-B bonds are presented in films with
different content of zirconium, which is peculiar to W$_2$B$_5$ and ZrB$_2$
phases.
No W-Zr bonds were found indicating the formation of solid solution as a result of
replacing the tungsten atoms by zirconium atoms in the W-Zr-B lattice.
Similar results were also obtained in W-Ti-B~\cite{32}, Ta-W-B~\cite{34} and
Zr-Ta-B~\cite{37} ternary TM boride films.
On the whole, the results of XRD, ED, and XPS studies and the data on lattice
parameter measurements allow us to make a conclusion that substitution solid solutions are
formed in ternary transition metal borides.

\section{Mechanical properties}

\paragraph{Hardness and Young's modulus}

The hardness of ternary transition metal borides was studied by various
techniques (nanoindentation and microindentation) in films deposited onto
different substrates (silicon single crystals, high-speed and stainless
steels, polycrystalline Al$_2$O$_3$, single crystalline sapphire).
Thus, the absolute values of the measured hardness and Young's modulus are
different depending on the film-substrate-method combination.
However, the analysis of publications showed that there are general
tendencies to a change in the mechanical properties in ternary transition
metal film borides.

The mechanical properties of the films in the Ti-W-B system were studied in
\cite{28,30,31,32}.
In \cite{28}, the effect of W addition on the hardness and Young's modulus
of TiB$_2$ films was studied.
The results obtained showed that those quantities steadily increased with
addition of tungsten in amounts of 3-to-10 at. \% due to the changes in the
microstructure of coatings from a typical columnar structure of TiB$_2$ to a
clear nano-composite structure of Ti-B-W (10 at. \%).
Similar dependence for Young's modulus of Ti-W-B films, as determined from
nanoindentation, was observed in \cite{30} with increasing W content in
the range of 4.5-34.6 at. \%.
However, a pronounced maximum of the hardness was reached for W contents of
33 at. \% and 43 at.\% on the metal sublattice. Using ab initio methods,
there were presented vacancies as a possible explanation for the stabilization of
the $\alpha$-phase in this ternary system.

\begin{figure}[htpb]
    \centering
    \includegraphics[width=0.5\linewidth]{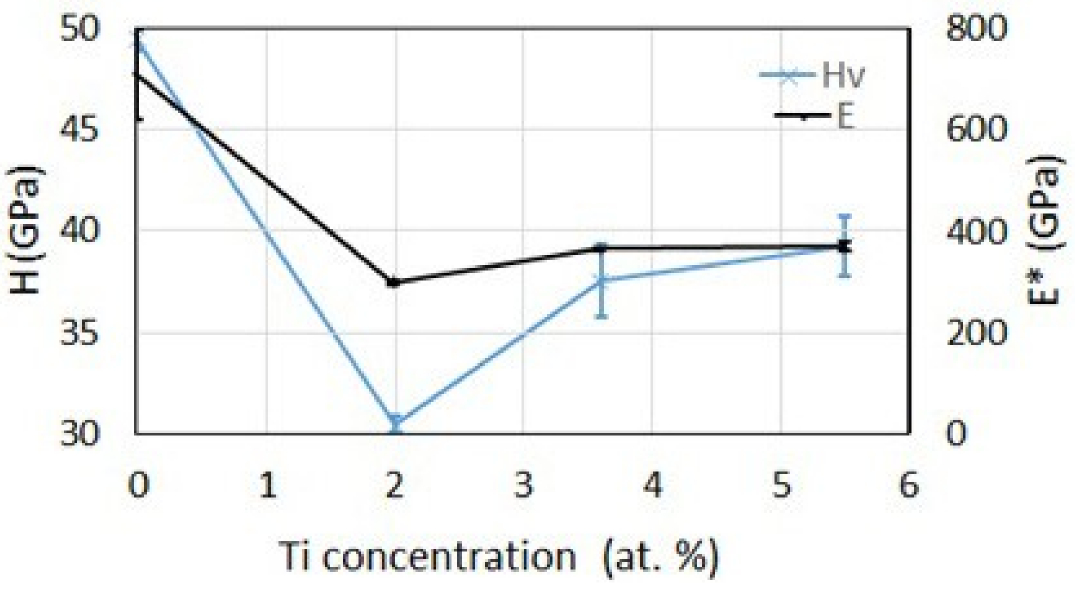}
    \caption{(Colour online) Hardness and effective elastic modulus of the deposited films
    as function of Ti concentration~\cite{32}.}
    \label{fig:19}
\end{figure}

The effect of Ti content on the hardness and Young's modulus of W-Ti-B films
was studied in \cite{31,32}.
Sobol et al.~\cite{31}  showed that these quantities steadily increase with
increasing Ti concentration from 0.6 at. \% to 10.2 at. \% as a result of
coating structure to become ordered on the basis of (Ti,W)B$_2$ phase.
Quite different result was obtained by Moscicki et al.~\cite{32}.
They showed that the deposited films without titanium exhibit properties as
presented in literature, namely: due to the crystalline columnar structure,
hardness reached 49.0$\pm$0.6 GPa with effective elastic modulus of
709.4$\pm$89.5 GPa (figure~\ref{fig:19}).
Addition of 2 at. \% Ti caused a two-fold decrease of the hardness and elastic
modulus.
This was a result of changing highly oriented crystalline structure to
amorphous phase.
The increasing of titanium content to C$_\mathrm{Ti}$ $\approx$ 3.6 at. \%
caused about 20 \% rise of the hardness and effective elastic modulus to
37.5$\pm$1.9 GPa and 341.2$\pm$2.4 GPa, respectively.
The highest hardness and elastic modulus values were measured for
C$_\mathrm{Ti}$ $\approx$ 5.5 at. \%.
The observed variation of these quantities was explained by structural changes
in the films first from crystalline to amorphous phase and next to hexagonal
$\alpha$-(W,Ti)B$_2$ structure at a higher content of titanium.

Similar behavior of the hardness and elastic modulus was observed in W-Zr-B
films~\cite{29}.
With increasing Zr content, the hardness and elastic modulus reduced slightly,
and then increased up to maximum values.
The decrease in mechanical properties of films with addition of Zr was
explained by structural changes in films.
At a low Ti content, an amorphization and, as a result, a decrease in compressive
stress occurred.
At higher Zr contents, the crystallization of a film resulted
in the formation of nanocomposite in which crystals of ZrB$_2$ are imbedded in
the Ti-Zr-B matrix.
The alloying of WB$_2$ with zirconium  also allows to change  the elastic
properties from brittle to ductile while maintaining the film superhardness and
incompressibility~\cite{29}.

\begin{figure}[htpb]
    \centering
    \includegraphics[width=0.45\linewidth]{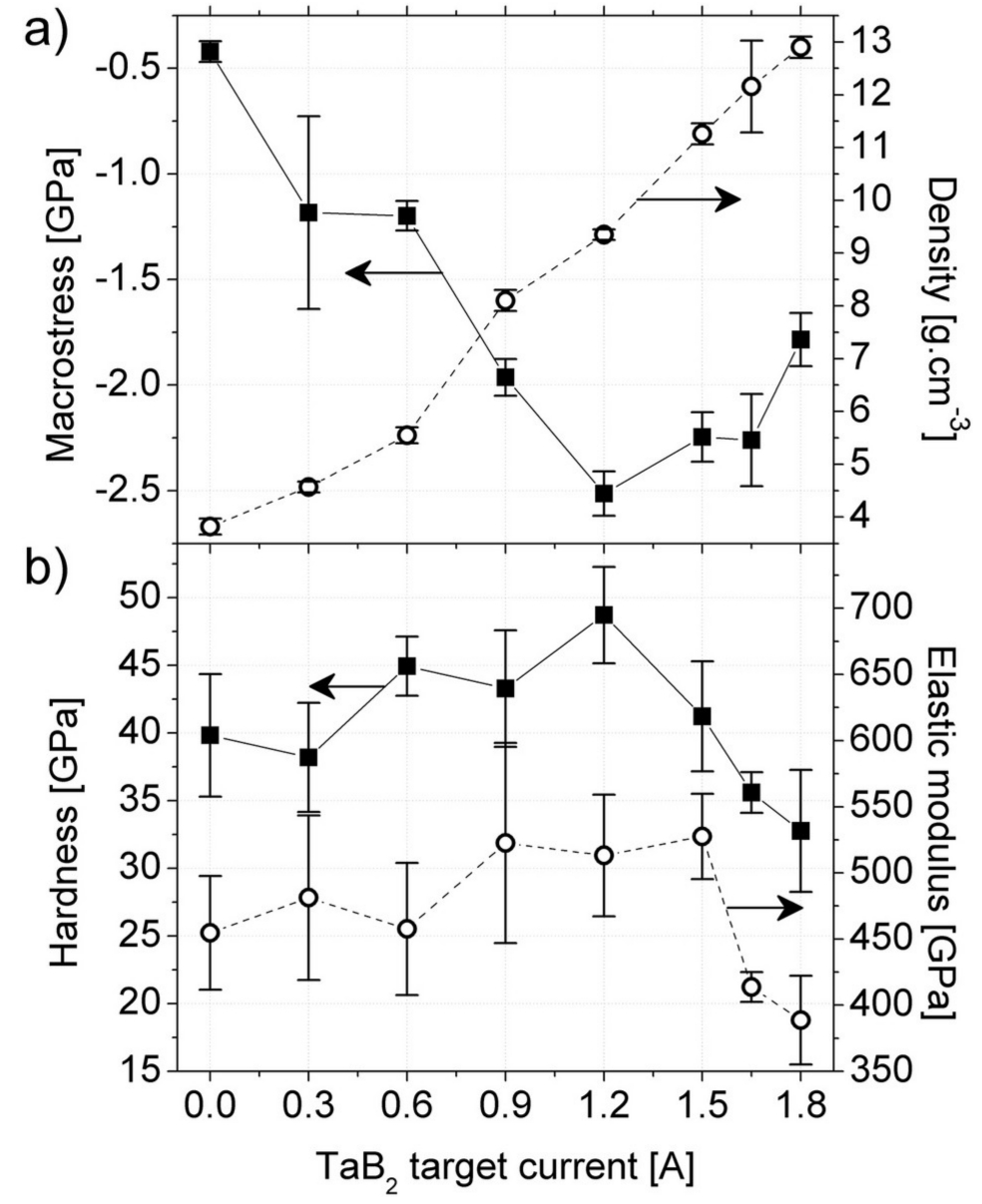}
    \caption{(a) Macrostress and density and (b) hardness and elastic modulus
    of co-deposited Ti-Ta-B films~\cite{36}.}
    \label{fig:20}
\end{figure}

In \cite{36}, the hardness and elastic modulus of Ti-Ta-B films were
investigated.
The variation of tantalum content in the films was attained by increasing the
sputtering current at the TaB$_2$ target.
By increasing the TaB$_2$ target current, a slight increase in hardness and
elastic modulus was observed up to stoichiometric Ti$_{0.18}$Ta$_{0.81}$B$_2$
coating.
By further increase of TaB$_2$ target current, a decrease of both mechanical
properties took place (figure~\ref{fig:20}).
The observed variations of hardness and elastic modulus were attributed to the
structure and stress measured in the deposited films.
It was shown that at low Ta content, the film structure exhibited a columnar
nature and enhanced the compressive macrostress.
The decrease of compressive macrostress in understoichiometric coatings
(B/Me < 2 in figure~\ref{fig:10}) is related to the transition from textured
nanocolumnar to amorphous structure.

A similar dependence of hardness and elastic modulus on the content of alloying
transition metal (Ta) was revealed for Zr-Ta-B films~\cite{37}.
With addition of Ta, the hardness and elastic modulus increased up to maximum
values, and then decreased.
The films with low content of Ta had a columnar structure in which thin B-rich
tissue phase, with strong covalent bonding, located at the column boundaries
which inhibits a column-boundary sliding.
An increase in mechanical properties was also attributed to the solid-solution
hardening effect.
The author of \cite{37}  also observed a decrease in the column width with
addition of Ta that also adds hardness to the film  via the Hall-Petch effect.
A further increase in Ta content resulted in the structure transformation towards
amorphous-like phase with grain boundaries enriched with metallic Ta, which
results in a decrease of hardness and elastic modulus of Zr-Ta-B films.

\paragraph{Strength properties}

The fracture toughness of transition metal borides was studied in
\cite{28,36,37} using nanoindentation technique.
In \cite{28}, the resistance to brittle cracking of Ti-W-B films was
investigated.
The analysis carried out according to the Laugier model showed that, as a
result of doping the TiB$_2$ coating with tungsten, there is a significant
increase in its fracture toughness.
For concentrations of 3-6 at.\% W, the fracture toughness of Ti-B-W coatings
reaches the value comparable to K$_\mathrm{IC}$ suitable for TiN and CrN
coatings.
The fracture toughness of Ti-B-W with 10 at. \% W is K$_\mathrm{IC}$ = 4.98
MPa$\cdot$m$^{1/2}$ and is nearly 7.5 times higher than for the TiB$_2$ coating
(K$_\mathrm{IC}$ = 0.67 MPa$\cdot$m$^{1/2}$).

Gran{\v c}i{\v c} et al.~\cite{36}  studied the fracture toughness of Ti-Ta-B
films.
It was shown that the coating structure and B/Me ratio appears to have an
important effect on the toughness of coatings. 
The coatings with B/Me > 2 exhibited a nanocolumnar structure and revealed
a lower toughness.
The under-stoichiometric (B/Me < 2) coatings with amorphous or nearly
amorphous structure showed an increased toughness.

Fracture toughness of Zr-Ta-B films was studied in \cite{37}.
It was revealed that with addition of Ta in Zr-B films, the toughness increased
from K$_\mathrm{IC} = 4.0$ MPa$\cdot$m$^{1/2}$ for ZrB$_{2.4}$ to 4.6
MPa$\cdot$m$^{1/2}$ and 5.2 MPa$\cdot$m$^{1/2}$ for
Zr$_{0.8}$Ta$_{0.2}$B$_{1.8}$ and Zr$_{0.7}$Ta$_{0.3}$B$_{1.5}$ film
compositions, respectively.
The nanostructures of these alloys consisted of a hard columnar phase with
metal-rich boundaries which inhibited the crack propagation while allowing
grain-boundary sliding under heavy loads.
Thus, Zr$_{0.8}$Ta$_{0.2}$B$_{1.8}$ and Zr$_{0.7}$Ta$_{0.3}$B$_{1.5}$ films
exhibited a dual hard/tough nature: the tough metal-rich phase at the boundaries
accommodates the ductility while the stiff nanosized columns provide high
hardness.

\paragraph{Tribology}

Wear resistance substantially depends on the properties that are inherent in
the films, namely: morphology, hardness, coefficient of friction, and
cohesion~\cite{41}.
The state of stress in the films is also important.
In this regard, compressive residual stresses are preferred, since tensile
stress is contributes to the cracking of the films under load.

\begin{figure}[htpb]
  \centering
  \includegraphics[width=0.7\linewidth]{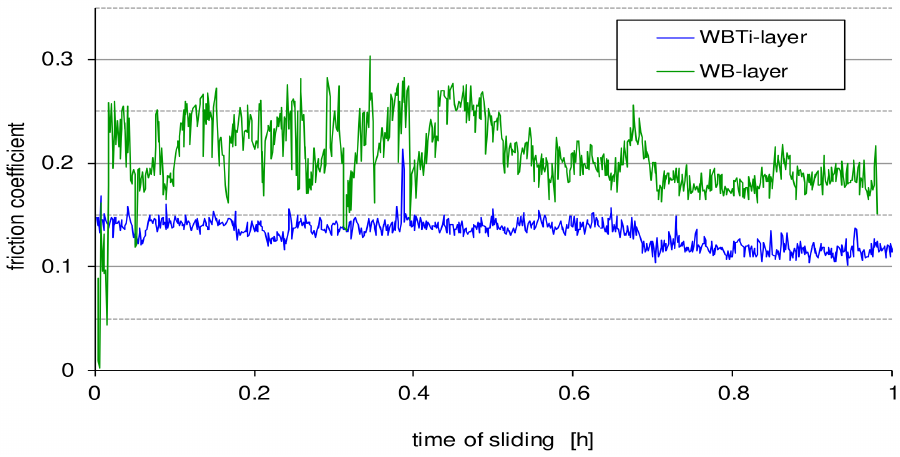}
  \caption{(Colour online) The time-dependent behavior of the friction coefficient of
  W-Ti(5.5 at. \%)-B and W-B coatings~\cite{32}.}
  \label{fig:21}
\end{figure}

Moscicki et al.~\cite{32}  studied the wear resistance of W-Ti-B films.
The comparative tribological test was conducted for of the W-B and W-Ti (5.5
at. \%)-B coatings.
The coating with titanium addition exhibited a higher wear resistance which
resulted from the structure of the compared layers.
Due to its compact structure, the titanium-alloyed film had a lower hardness,
although the crack resistance was higher.
The non-columnar structure contained fewer defects, which made the film more
ductile and resistant to cracking.
The friction coefficient was lower for the W-Ti-B film than for the W-B film
(figure~\ref{fig:21}).

\paragraph{Thermal stability}
%\subsection{Thermal stability}

Euchner et al.~\cite{30} studied the effect of annealing on the structure and
properties of W-doped (4.5 to 34.6 at. \% W) TiB$_2$ films.
\textcolor{black}{They showed that these films} exhibited no changes in their
XRD patterns upon 30-min vacuum annealing at 1000\textcelsius{} suggesting an
outstanding thermal stability of a dense growth morphology.
A similar result was obtained in a vacuum annealed W-Ta(7--26 at. \%)-B
films~\cite{33}.
All compositions remained in their single-phased $\alpha$-structure up to
800\textcelsius{}.
At 1000\textcelsius{}, the binary metastable $\alpha$-WB$_{2-z}$ decomposed to form
t-WB and $\omega$-W$_2$B$_{5-z}$.
The addition of just 7 at. \% Ta to the metal sublattice postponed the
decomposition and transformation of the $\alpha$-structure to 1200\textcelsius{}.
Further alloying of Ta even increased the decomposition temperature and
ensured the thin film materials for Ta contents above 14 at. \% to be
single-phased $\alpha$-structured even up to 1200\textcelsius{}.
These results prove the concept of stabilizing phases due to the addition of
an appropriate alloying element.

\paragraph{Oxidation and corrosion resistance}

\begin{figure}[htpb]
  \centering
  \includegraphics[width=0.6\linewidth]{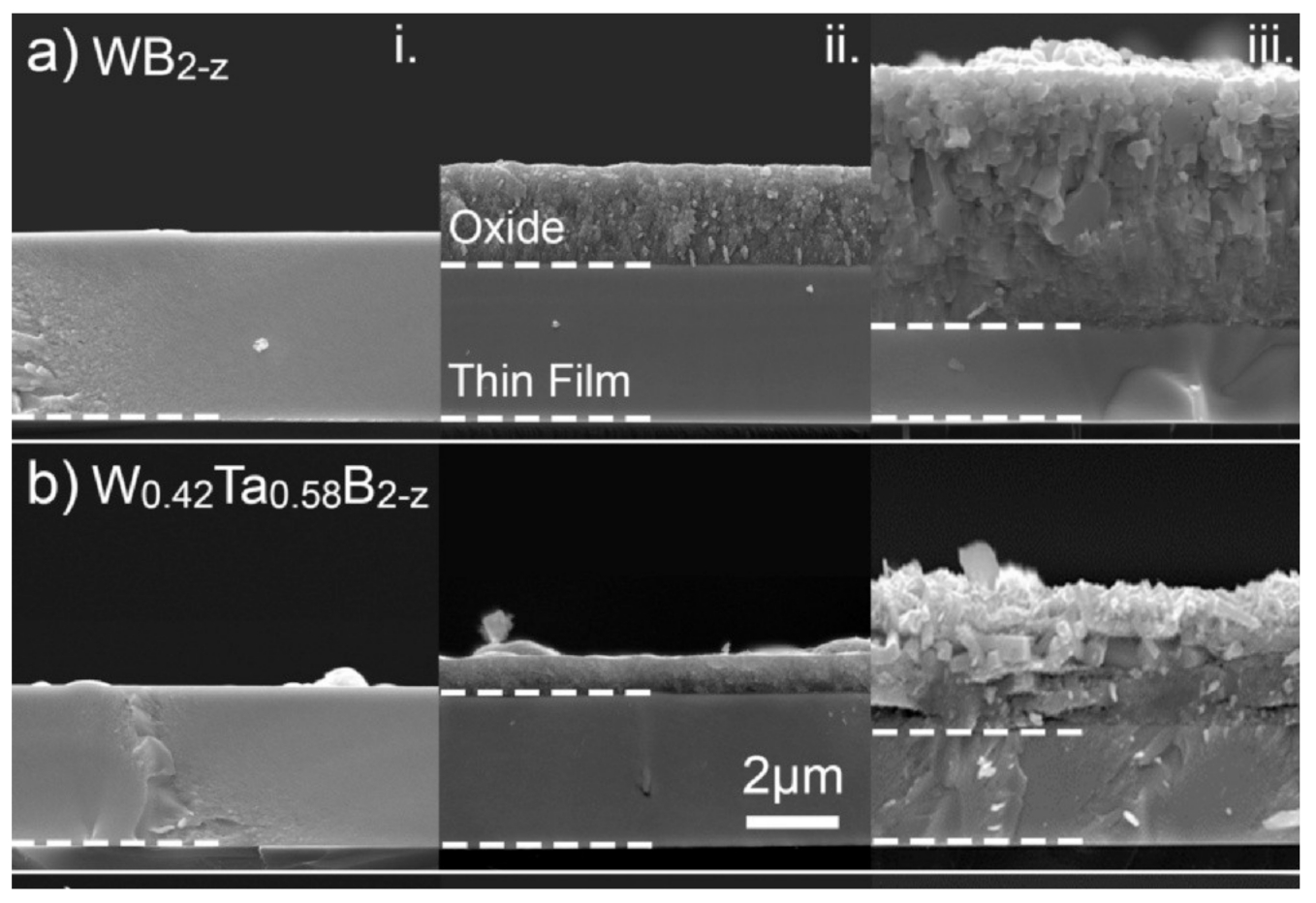}
  \caption{Cross sectional micrographs of (a) WB$_{2-z}$ and
    (b) W$_{0.42}$Ta$_{0.58}$B$_{2-z}$ oxidized films.
    Sections (a), (b) show selected coatings in (i) as deposited state,
    (ii) after 100 min of oxidation at 600\textcelsius{} and (iii) after 100 min
    of oxidation at 700\textcelsius{}.
    The dashed lines are highlighting  the interface between substrate and
    film as well as between film and oxide scale (from bottom to
  top)~\cite{34}.}
  \label{fig:22}
\end{figure}

Oxidation behavior in the ambient air of Ta-W-B films was investigated
in~\cite{34}.
The compositions in the full range from WB$_{2-z}$,
W$_{0.85}$Ta$_{0.15}$B$_{2-z}$, W$_{0.66}$Ta$_{0.34}$B$_{2-z}$,
W$_{0.42}$Ta$_{0.58}$B$_{2-z}$, W$_{0.19}$Ta$_{0.81}$B$_{2-z}$, to
TaB$_{2-z}$ were studied.
The oxidation tests in the ambient air at 500, 600, 700\textcelsius{} for 1, 10, 100
and 1000 min showed a decrease in the oxide layer thickness with increasing Ta
content for all temperatures applied (figure~\ref{fig:22}).
Generally, the addition of Ta to WB$_{2-z}$ based coatings retarded
the oxide scale kinetics through the formation of denser, less volatile,
and adherent scales, which is a key factor for the enhanced oxidation
resistance.
An optimum composition of
W$_{1-x}$Ta$_x$B$_{2-z}$ coatings would be in the range of $x = 0.2-0.3$,
combining the enhanced fracture toughness ($\geqslant$ 3.0 MPa$\cdot$m$^{1/2}$) and the super hardness ($\geqslant$ 40 GPa) next to a decent oxidation resistance.

Corrosion resistance of W-Ti-B films placed onto the plates of 304 stainless
steels (SS 304) was investigated in the electrochemical corrosion tests ---
impedance spectroscopy and polarization --- in 0.5M NaCl solution~\cite{32}.
As it was shown by structure study, the addition of titanium to the WB$_2$
caused the Ti-doped films to become denser and to lose their columnar structure.
As a result, less direct diffusion paths for the corrosive medium were formed
which resulted in the improvement of corrosion resistance.
The best protective efficiency was achieved for
W$_{0.92}$Ti$_{0.08}$B$_{4.5}$ coating.
The protective efficiency decreased with the rise of titanium content and the
associated increase in hardness.

The effect of boron on the corrosion resistance of Ni-Cr-B coatings deposited
on commercial steel SA213-T2 by spraying NiCrB cored wires was 
also investigated in the electrochemical corrosion tests~\cite{38}.
The hot corrosion tests were performed at 800\textcelsius{} in molten mixed salt
Na$_2$SO$_4$ + 10 wt. \% NaCl for 200h with duration of 10h to evaluate the
corrosion behavior of the coatings.
The results obtained showed that boron promoted a successive and thin
protective oxide scale on the surface of NiCrB coatings.
The coatings with B content 0.5--3 wt. \% had a much lower weight gain in the
hot corrosion test.
However, too much boron was harmful to hot corrosion property due to the
brittle borides and stress-inducing cracks, which worked as the path tunnel
of the corrosion substance.

\color{black}
\paragraph{Future prospects for research}

The research of metal borides is a very old field, and ever since the structure and wide
range of metal diborides have been studied exhaustively.
Therefore, it is necessary to move to ternary and higher borides as starting
materials.
These compositions are all the more challenging due to the factors such as
phase stability relative to metal substitutions and simultaneous equilibrium
of several phases in the product.
The vast array of determined crystal structures to date offers a potential
for targeted approaches to the prediction of structure and properties.

Supersaturated solid solutions of many ternary borides have been studied here,
which are based on binary constituents that in principle prefer to
crystallize in the same or different structural modifications.
It was shown that such ternary borides represent a new class of materials
having useful properties.
The results obtained for ternary borides open the prospects for developing
new materials based on combination of metal diborides in quaternary
structures or even high entropy alloys.
The design of such alloys may  not be so
difficult if their
approximate optimal composition is known.
The latter is quite possible when the results of first principles
calculations are available.
Therefore, we foresee that the combination of experiment and first principles
calculations should be a particularly successful approach in the development
of the films based on complex high-entropy alloys.
\color{black}

\section{Conclusions}

\color{black}
The review of recent achievements in the bulk and film materials based on
ternary transition metal diboride alloys is given.
First principles investigations of the structure and properties of the
M$^\mathrm{I}_{1-x}$M$^\mathrm{II}_x$B$_2$ and
M$^\mathrm{I}_{x}$M$^\mathrm{II}$B$_x$
($x=1$ or 2, M$^\mathrm{I}$ and M$^\mathrm{II}$ are transition metals)
ternary borides showed that they possess high hardness, and hence high
fracture toughness and exhibit ductility comparable to the ductility of WC
(except for Ti$_{1-x}$Zr$_x$B$_2$ alloys that are highly brittle materials).
These characteristics and the capability of controlling the properties by changing
the composition make the ternary boride alloys suitable for many
applications.
Among them is the use of these alloys as the main components in
wear-resistant and protective coatings resistant to extremely high
temperatures.
\color{black}

The ternary boride films based on transition metal diborides can form solid
solution single-phase AlB$_2$-type or WB$_2$-type structures through partial
replacing of the matrix atoms by the added atoms of another transition metal.
The formation of solid solutions is experimentally accessible over a large
composition range.
The films exhibit a dense columnar or near-amorphous structure which is
stable upon annealing up to 1200\textcelsius{}.
Ternary transition metal borides show a fairly high level of hardness,
fracture toughness, good tribological properties.
Therefore, the films of ternary transition metal borides have good prospects as
materials used as protective coatings in various fields of technology.

%\bibliographystyle{cmpj}
%\bibliography{manuscript}

% If you have problems with typesetting in ukrainian uncomment lines below.

\ukrainianpart
\title{Структура та властивості плівок на основі боридів потрійних перехідних металів:
	теорія та експеримент}
\author{А. А. Онопрієнко, В. І. Іващенко, В. I. Шевченко%
	}
\address{Інститут проблем матеріалознавства ім. І. М. Францевича НАН України, Київ, Україна}

\makeukrtitle
\begin{abstract}
	\tolerance=3000
	В огляді приведено результати теоретичних та експериментальних досліджень структури,
	зв’язку між атомами, механічних властивостей, термічної стабільності, стійкості до
	окислення та корозії плівок на основі боридів потрійних перехідних металів.
	
	\keywords потрійні бориди перехідних металів, плівки, структура, властивості
\end{abstract}

\end{document}